\documentclass[a4paper,11pt]{article}
\usepackage{jheppub}
\usepackage{subfig}
\usepackage{xcolor}
\usepackage{array}

\newcommand{\CLASS}{\texttt{CLASS}}

\title{Probing Confining Dark Sectors with Cosmological Perturbations}

\author[a]{Daven Wei Ren Ho,}
\author[b,c]{Ameen Ismail,}
\author[a]{Yuhsin Tsai}

\affiliation[a]{Department of Physics and Astronomy, University of Notre Dame, Notre Dame, IN 46556, USA}
\affiliation[b]{Enrico Fermi Institute, University of Chicago, Chicago, IL 60637, USA}
\affiliation[c]{Leinweber Institute for Theoretical Physics, University of Chicago, Chicago, IL 60637, USA}

\emailAdd{dho2@nd.edu}
\emailAdd{ameenismail@uchicago.edu}
\emailAdd{ytsai3@nd.edu}

\abstract{
Dark matter may emerge as a composite state of a dark sector which confines in a strongly first-order phase transition (PT). To avoid structure formation constraints on warm dark matter, the dark PT must occur above the keV scale. We investigate the cosmological signatures of this scenario, focusing on a keV- to MeV-scale PT. The stochastic nature of bubble nucleation sources curvature perturbations that can be constrained by various cosmological observations. Composite dark matter inherits the isocurvature perturbations generated during the PT and sources large-scale curvature perturbations. In contrast to a PT that reheats into dark radiation, the slower redshifting of dark matter enhances the infrared tail of the curvature perturbation upon horizon entry. The PT-induced perturbations compete with the suppression of the matter power spectrum due to the free-streaming of composite dark matter. We place limits on the PT strength and temperature from cosmic microwave background anisotropies, the Lyman-$\alpha$ forest, and other probes of the small-scale matter power spectrum. In a minimal scenario where the relic density is determined by the PT parameters, this provides a concrete example of a dark matter model that is testable via measurements of cosmological perturbations---even in the absence of a sizable nongravitational coupling to the visible sector.
}

\begin{document} 
\maketitle
\flushbottom

%%%%%%%%%%%%%%%%%%%%%%%%%%%%%%%%%%%%%%%%%%%%%%%%%%%%%%%
%%%%%%%%%%%%%%%%%%%%%%%%%%%%%%%%%%%%%%%%%%%%%%%%%%%%%%%
\section{Introduction}
%%%%%%%%%%%%%%%%%%%%%%%%%%%%%%%%%%%%%%%%%%%%%%%%%%%%%%%
%%%%%%%%%%%%%%%%%%%%%%%%%%%%%%%%%%%%%%%%%%%%%%%%%%%%%%%

The microscopic nature of dark matter (DM) is a perennial mystery in particle physics. The DM may even be part of a larger dark sector (DS) involving multiple fields that do not carry any Standard Model (SM) gauge charges. A DS can exhibit rich structure and phenomenology that would be overlooked if one focused only on simpler models (see e.g.~\cite{Harris:2022vnx,Krnjaic:2022ozp} for reviews).

In this work we are interested in the possibility that a DS undergoes a strongly first-order phase transition (PT).
One compelling way this may arise is if the DM is a composite of a strongly-coupled, nearly conformal DS.
There are several reasons one might consider such a scenario.
In composite Higgs~\cite{Bellazzini:2014yua,Panico:2015jxa,Contino:2003ve,Agashe:2004rs} or Randall--Sundrum models~\cite{Randall:1999ee,Randall:1999vf} (related through the anti-de Sitter / conformal field theory (AdS/CFT) correspondence~\cite{Rattazzi:2000hs,Arkani-Hamed:2000ijo}), which were initially developed to address the hierarchy problem, some of the SM degrees of freedom are composites of a near-conformal DS. This motivates models in which the DM also arises from such a DS~\cite{Agashe:2004ci,Agashe:2004bm,Agashe:2007jb,Agashe:2009ja,McDonald:2010iq,Medina:2011qc,Kim:2016jbz,Folgado:2019sgz,Donini:2025cpl,Donini:2025qrf}.
Various works have also considered DM arising from a more general conformal DS, not necessarily related to the hierarchy problem~\cite{Bai:2009ms,McDonald:2010fe,McDonald:2012nc,vonHarling:2012sz,Robinson:2012wu,Robinson:2014bma,Blum:2014jca,Brax:2019koq,Bernal:2020fvw,Fuks:2020tam,Redi:2020ffc,Bernal:2020yqg,deGiorgi:2021xvm,deGiorgi:2022yha,Ahmed:2023vdb,Ferrante:2023bcz,Ahmed:2025ldh}.
Particularly relevant for our purposes is the conformal freeze-in (COFI) framework, in which feeble interactions with the SM populate a conformal DS~\cite{Hong:2019nwd,Hong:2022gzo,Chiu:2022bni,Hong:2024zsn,Luo:2025alo}.

We expect that the conformal sector is in the hot, deconfined phase initially, and it undergoes a PT in the early Universe. CFT stuff confines into the bound states of the cold phase, including the DM candidate.
The dynamics of the conformal PT has been a subject of intense study, largely motivated by the possibility of observing a resulting gravitational wave signal at future experiments, e.g.~\cite{Creminelli:2001th,Randall:2006py,Bunk:2017fic,vonHarling:2017yew,Baratella:2018pxi,Agashe:2019lhy,Agashe:2020lfz,Csaki:2023pwy,Mishra:2023kiu,Eroncel:2023uqf,Mishra:2024ehr,Mishra:2026lvq,Ismail:2026tyb}.
The PT is generically first-order; in some cases it can be very strong, with an inverse duration $\beta/H_{\rm PT} \lesssim 10$.

In recent years, it has been recognized that strong first-order PT at late times can leave observable imprints on the cosmic microwave background (CMB) and structure formation signals~\cite{Liu:2022lvz,Elor:2023xbz,Buckley:2024nen,Greene:2024xgq,Buckley:2025zgh,Xu:2025zsv,Koren:2025ymq,Chang:2025uvx,Zebrowski:2026pye,Greene:2026gnw}. The key idea is that such transitions proceed via stochastic bubble nucleation, so the completion time varies slightly across different patches of the Universe. This inhomogeneity sources both scalar and tensor mode perturbations, producing power spectra with characteristic peaks at a scale set by the typical bubble size at the end of the PT. At the same time, a variety of promising proposals have emerged for probing the matter power spectrum on sub-Mpc scales~\cite{Chluba:2012we,Cyr:2023pgw,Graham:2024hah,Qin:2025ymc,Bringmann:2025cht,Lee:2020wfn}. This makes it timely to systematically explore the possibility of testing low-scale DS through the cosmological signatures of their PTs. That is the aim of this work. While our analysis is partly motivated by conformal DS models, the results apply more generally.

In this work, we focus on scalar perturbation signals from a confining DS undergoing a strongly first-order PT at the keV–MeV scale. We introduce a phenomenological framework in Section~\ref{sec:model}, in which the DS is completely decoupled from the visible sector. We first compute the DM relic abundance assuming that all latent heat from the PT is transferred into DM production, directly tying the relic density to the PT parameters. In this sense, experimental constraints on the PT can, in principle, be used to assess the viability of this DM scenario—even in the absence of any nongravitational couplings to the SM, a ``nightmare scenario". We then extend the analysis to cases where a fraction of the latent heat is converted into dark radiation (DR), which allows the observed DM abundance to be reproduced for stronger PTs and leads to a clearer interpretation of the resulting constraints.

We calculate the impact of the PT on the matter power spectrum in Section~\ref{sec:perturbations}, building on the framework of~\cite{Elor:2023xbz,Greene:2026gnw}. Our analysis contains two main new ingredients. First, we identify a $1/k^2$ enhancement of the curvature power spectrum on large scales (small-$k$ modes) relative to the case where the DS consists only of DR. This enhancement arises because, once the DS begins redshifting as matter, its energy-density fraction grows linearly with the scale factor until matter-radiation equality. Second, for the PT temperatures considered here, the composite DM is naturally produced warm. The resulting free-streaming suppresses the matter power spectrum through the transfer function, while the enhanced PT-induced perturbations partially compensate for this suppression, leading to a modest relaxation of the conventional WDM mass bound.

In Section~\ref{sec:analysis}, we analyze constraints on the power spectrum resulting from the PT, using a range of CMB and structure formation observations. For PTs occurring at temperatures in the $1$--$10$ keV range, where the generated WDM perturbations are most significant, we perform dedicated likelihood studies using CMB anisotropy and Lyman-$\alpha$ forest data. More specifically, we obtain CMB and BAO constraints using a Markov Chain Monte Carlo (MCMC) likelihood analysis; we estimate the Lyman-$\alpha$ bound using a two-parameter fit of the predicted matter power spectrum to compressed data from the eBOSS flux power spectrum~\cite{Bird:2023,Fernandez:2024,He:2025}. We also obtain bounds by comparing our predicted power spectrum to published bounds in the literature, including constraints from CMB spectral distortions~\cite{Chluba:2012we,Cyr:2023pgw}, dynamical heating of stars in ultra-faint dwarfs~\cite{Graham:2024hah}, early reionization~\cite{Qin:2025ymc}, formation of ultracompact minihalos~\cite{Bringmann:2025cht}, and pulsar timing arrays~\cite{Lee:2020wfn}. 

Using these results, we derive current and projected constraints on the PT parameters in Section~\ref{sec:results}. In the scenario where the PT predominantly reheats into composite DM, we identify a viable region with PT temperatures and DM masses in the $1$--$10$~keV range. In this regime, the most sensitive probes arise from measurements of the matter power spectrum through the CMB and Lyman-$\alpha$ forest. We also consider the scenario in which a substantial fraction of the latent heat is converted into DR. In this case, we place constraints on PT temperatures spanning the keV–MeV range, with the strongest limits coming from recently proposed probes of the power spectrum on sub-Mpc scales. We summarize the results in Section~\ref{sec:conclusions}.

%%%%%%%%%%%%%%%%%%%%%%%%%%%%%%%%%%%%%%%%%%%%%%%%%%%%%%%
%%%%%%%%%%%%%%%%%%%%%%%%%%%%%%%%%%%%%%%%%%%%%%%%%%%%%%%
\section{Thermal history of the confining DS}\label{sec:model}
%%%%%%%%%%%%%%%%%%%%%%%%%%%%%%%%%%%%%%%%%%%%%%%%%%%%%%%
%%%%%%%%%%%%%%%%%%%%%%%%%%%%%%%%%%%%%%%%%%%%%%%%%%%%%%%

%\subsection{Models}
We consider a DS that undergoes a first-order PT and confines in the early universe. We assume the DS interacts sufficiently weakly with the SM sector such that it never thermalizes with the SM bath. The DS can be populated through mechanisms such as freeze-in or asymmetric reheating, and therefore has its own temperature $T_d$, distinct from the SM photon temperature.

As a concrete example, one may consider a near-conformal DS that is spontaneously broken and confines at low temperatures. If the DS is populated via freeze-in, this corresponds essentially to the COFI framework~\cite{Hong:2019nwd,Hong:2022gzo,Chiu:2022bni,Hong:2024zsn,Luo:2025alo}. 

Let $T_{\rm PT}$ denote the temperature of the SM bath at the time of the dark PT. We make the following assumptions about the PT:
\begin{itemize}
\item The PT is strongly first-order and supercooled, with the DS dominated by its vacuum energy $\rho_{\rm vac}$ prior to the transition.
\item The latent heat $\rho_{\rm vac}$ released during the PT is fully converted into DS composites and DR.
\end{itemize}

Since we are interested in scenarios where DM is composed of DS states, at least one composite state must be stable on cosmological timescales. The DM relic abundance can then be set by various decay or annihilation processes among the composite DS states. A minimal and predictive possibility is that essentially all of the latent heat released during the PT is transferred into the DM energy density, with any remaining DS species efficiently annihilating or decaying into the stable DM state.

Additional dynamics in the DS, such as a freeze-out mechanism, could further deplete the DM abundance (some possible mechanisms were discussed in~\cite{Freese:2023fcr}). However, provided that the DS and the SM are not in thermal contact, the relic abundance cannot be made larger. That is, this minimal scenario could be interpreted as an upper bound on the relic abundance. In what follows we will also consider an example of this. Specifically, we will study the case where only a fraction $f_{\rm DM}$ of the latent heat contributes to the DM relic density, with the remainder becoming DR.

In summary, we consider two DS reheating scenarios arising from the first-order PT:
\begin{itemize}
    \item \textbf{DM-only}: All the latent heat of the PT contributes to the DM energy density. The DM may be warm, which is discussed further in Section~\ref{sec:perturbations}. 
    \item \textbf{DM+DR}: A fraction of the latent heat contributes to the DM matter density, with the rest becoming DR. We assume the DM is cold in this case.
\end{itemize}
\vspace{1em}
{\bf DM-only:} We define the PT strength as $\alpha_{\rm PT} \equiv \rho_{\rm vac} / \rho_{\rm SM}(T_{\rm PT})$, where $T_{\rm PT}$ denotes the temperature of the SM sector at the time the dark PT occurs. Focusing on PTs that occur deep in the radiation-dominated era, the energy density in the dark sector immediately after the PT is given by
\begin{equation}\label{eq:rhod1}
    \rho_d(T_{\rm PT}) = \alpha_{\rm PT} \,\frac{\pi^2}{30} g_*(T_{\rm PT})\,T_{\rm PT}^4\,.
\end{equation}

Initially, the DS temperature can be much larger than the mass of composite DM, $T_{\rm PT}\gg m_{\rm DM}$. During this epoch, $\rho_d$ redshifts as radiation, and the DS temperature is determined by $\rho_d = A T_d^4$. Here $A$ is an order-one, model-dependent constant related to the effective number of relativistic degrees of freedom in the DS; the value of $A$ does not affect any of our results. The dark sector energy density and temperature depend on the SM temperature $T$ as
\begin{align}\label{eq:rhod2}
    \rho_d(T) = \frac{\pi^2}{30}\alpha_{\rm PT}\, {g_*(T) \,T^4}\,,\qquad 
    T_d(T) = \left[ \frac{\pi^2\alpha_{\rm PT}}{30 A} \right]^{1/4} {g_*(T)^{1/4} \,T}\,. 
    %= T_d(T_{\rm PT}) \frac{g_*(T)^{1/4} T}{g_*(T_{\rm PT})^{1/4} T_{\rm PT}} .
\end{align}

This epoch ends when the dark sector temperature is comparable to the DM mass. Let $T_m$ denote the temperature of the SM sector when $T_d=m_{\rm DM}$. From Eqs.~\eqref{eq:rhod1} and \eqref{eq:rhod2}, we have
\begin{equation}\label{eq:tm}
    m_{\rm DM} = T_d(T_m) = T_d(T_{\rm PT}) \frac{g_*(T_m)^{1/4} T_m}{g_*(T_{\rm PT})^{1/4} T_{\rm PT}} ,
\end{equation}
while the energy density in the DS is
\begin{equation}
    \rho_m \equiv \rho_d(T_m) = \rho_d(T_{\rm PT})\left[ \frac{g_*(T_m) T_m^4}{g_*(T_{\rm PT}) T_{\rm PT}^4} \right].
\end{equation}
Assuming the yield of composite DM has frozen out, its energy density subsequently redshifts as cold matter, resulting in a present-day energy density of
\begin{equation}
    \rho_d(T_0) = \frac{\pi^2}{30}\alpha_{\rm PT} g_*(T_{\rm PT}) T_{\rm PT}^4 \left[ \frac{g_*(T_m) T_m^4}{g_*(T_{\rm PT}) T_{\rm PT}^4} \right] \left[ \frac{g_*(T_0) T_0^3}{g_*(T_m) T_m^3} \right] = \frac{\pi^2}{30} \alpha_{\rm PT} g_*(T_0) T_m T_0^3\,,
\end{equation}
and the relic abundance
\begin{equation}\label{eq:relicabundance}
    \Omega_{\rm DM} h^2 = 0.12 \,r \left(\frac{\alpha_{\rm PT}}{6.7 \times 10^{-4}}\right)\left( \frac{T_{\rm PT}}{1 \rm ~keV} \right)\,.
\end{equation}
Here we define $r\equiv T_m / T_{\rm PT}$. Notice that for a given $T_{\rm PT}$, $r$ cannot be arbitrarily small since one requires $T_m \gtrsim 1~\mathrm{keV}$ to evade WDM constraints.
\\
\\
{\bf DM+DR:} In the case where a fraction $f_{\rm DM}$ of the latent heat contributes to the DM relic abundance, the relic abundance is just Eq.~\eqref{eq:relicabundance} with the replacement $\alpha_{\rm PT} \rightarrow f_{\rm DM} \alpha_{\rm PT}$. In this scenario, we assume that the DM becomes cold and nonrelativistic right after the PT, so its energy density always redshifts as matter. The relic abundance is then given by
\begin{equation}
    \Omega_{\rm DM} h^2 = 0.12 \,f_{\rm DM}\left(\frac{\alpha_{\rm PT}}{6.7 \times 10^{-4}} \right)\left(\frac{T_{\rm PT}}{1 \rm ~keV}\right)\,.
\end{equation}
By adjusting $f_{\rm DM}$, one can reproduce the observed relic abundance for a given $\alpha_{\rm PT}$. Although this additional assumption reduces the generality of the DS model, it provides a clearer interpretation of the cosmological constraints by avoiding an overproduction of DM.

%%%%%%%%%%%%%%%%%%%%%%%%%%%%%%%%%%%%%%%%%%%%%%%%%%%%%%%
%%%%%%%%%%%%%%%%%%%%%%%%%%%%%%%%%%%%%%%%%%%%%%%%%%%%%%%
\section{Curvature perturbations from a  dark-confinement PT}\label{sec:perturbations}
%%%%%%%%%%%%%%%%%%%%%%%%%%%%%%%%%%%%%%%%%%%%%%%%%%%%%%%
%%%%%%%%%%%%%%%%%%%%%%%%%%%%%%%%%%%%%%%%%%%%%%%%%%%%%%%

In this section we study the curvature perturbations arising from the dark PT. We focus on the DM-only scenario discussed above; in the DM+DR scenario,  most of the latent heat of the PT is converted into DR, so the curvature perturbations are similar to those derived in~\cite{Elor:2023xbz}.

The PT proceeds via bubble nucleation, and due to its stochastic nature, the completion time $t_c$ varies across different regions of the universe. In~\cite{Elor:2023xbz}, the authors computed the two-point function of the deviation of $t_c$ from its spatial average, $\delta t$. %Although our CMB and Lyman-$\alpha$ analyses use the full numerical spectrum from the integral expression in~\cite{Elor:2023xbz}, \dho{Note: for the CMB and Lyman-alpha analysis, I used the broken power law form of the spectrum (c.f. start of section 3.3) but only the superhorizon slope matters for the $T_{\rm PT}$ range considered} 
The result can be well approximated by a broken power law, providing a simpler estimate of cosmological bounds that are primarily sensitive to large-scale perturbations:
\begin{equation}\label{eq:Pdt}
    \mathcal{P}_{\delta t} \approx
    \begin{cases}
        3(8\pi) v_w^3\left( \frac{\beta}{H_{\rm PT}} \right)^{-2} \left( \frac{k}{a_{\rm PT} \beta} \right)^3\,,  & k/a_{\rm PT} \ll \beta\,, \\
        0.7 \left( \frac{\beta}{H_{\rm PT}} \right)^{-2} \left( \frac{k}{a_{\rm PT} \beta} \right)^{-3}\,, & k/a_{\rm PT}\, \gg \beta\,.
    \end{cases}
\end{equation}
Here $\beta$ denotes the inverse duration of the PT in physical time, while $a_{\rm PT}$ and $H_{\rm PT}$ are the scale factor and Hubble parameter at the time of the transition, respectively. We assume runaway bubbles with wall velocity $v_w = 1$, as is typical in strongly supercooled conformal PTs~\cite{Bigazzi:2021ucw,Bea:2021zsu}. Note that the regime $k/a_{\rm PT} \gtrsim \beta$ is sensitive to turbulence and hydrodynamic effects, and we therefore omit this part of the spectrum in our constraint analysis.

\subsection{Initial perturbation before horizon entry}

The isocurvature perturbations in the DS generated by the PT eventually source curvature perturbations $\zeta$. Consider two patches $A$ and $B$ in which the PT completes at slightly different times, $t_c^{A}$ and $t_c^{B}$. Immediately after the phase transition, the energy density in both patches is $\rho_d^{A,B}(t_c^{A,B}) = \rho_{\rm vac}$. The energy density subsequently redshifts as radiation until $\rho_d = \rho_m$. The corresponding time at which this occurs in each patch, $t_m^{A,B}$, is then given by
\begin{equation}
    t_m^{A,B} = \sqrt{ \frac{\rho_{\rm vac}}{\rho_m}}\, t_c^{A,B} .
\end{equation}
We remark that since $t_m \propto t_c$, their perturbations are related as $\delta t_m / t_m = \delta t_c / t_c$. Here we only work to first order in the perturbation $\delta t_c = t_c^{B} - t_c^{A}$.

The energy density in each patch at some later time $t > t_c$ is
\begin{equation}
    \rho_d^{A,B}(t) =
    \begin{cases}
        \rho_{\rm vac} \left( \frac{t_c^{A,B}}{t} \right)^2 & t < t_m^{A,B} \\
        \rho_{\rm vac} \left( \frac{t_c^{A,B}}{t_m^{A,B}} \right)^2 \left( \frac{t_m^{A,B}}{t} \right)^{3/2} & t > t_m^{A,B} .
    \end{cases}
\end{equation}
The density perturbation is therefore 
\begin{equation}
    \frac{\delta \rho_d}{\rho_d}  =
    \begin{cases}
        2 \frac{\delta t_c}{t_c}  & t < t_m \\
        \frac{3}{2} \frac{\delta t_c}{ t_c}  & t > t_m .
    \end{cases}
\end{equation}
Finally, we estimate the curvature perturbation in the spatially-flat gauge, neglecting the adiabatic perturbation,
\begin{align}\label{eq:curvature}
    \zeta &\approx -\frac{H \delta \rho_d}{\dot{\rho}_{\rm SM}}= \frac{\alpha_{\rm PT}}{2} \frac{\delta t_c}{t_c} F(\sqrt{t/t_m}) \,,\quad 
    F(x) =
    \begin{cases}
        1 & x < 1 \\
        3x/4 & x > 1
    \end{cases} .
\end{align}
Here we have used the fact that $\rho_d \ll \rho_{\rm SM}$, as required by cosmological constraints. The discontinuity between the relativistic ($t < t_m$) and nonrelativistic ($t > t_m$) regimes arises from the overly simplified treatment of the DS energy evolution. We will shortly provide a more careful calculation in which the perturbation remains continuous between different scales.

\begin{figure}
    \centering
    \includegraphics[width=0.8\textwidth]{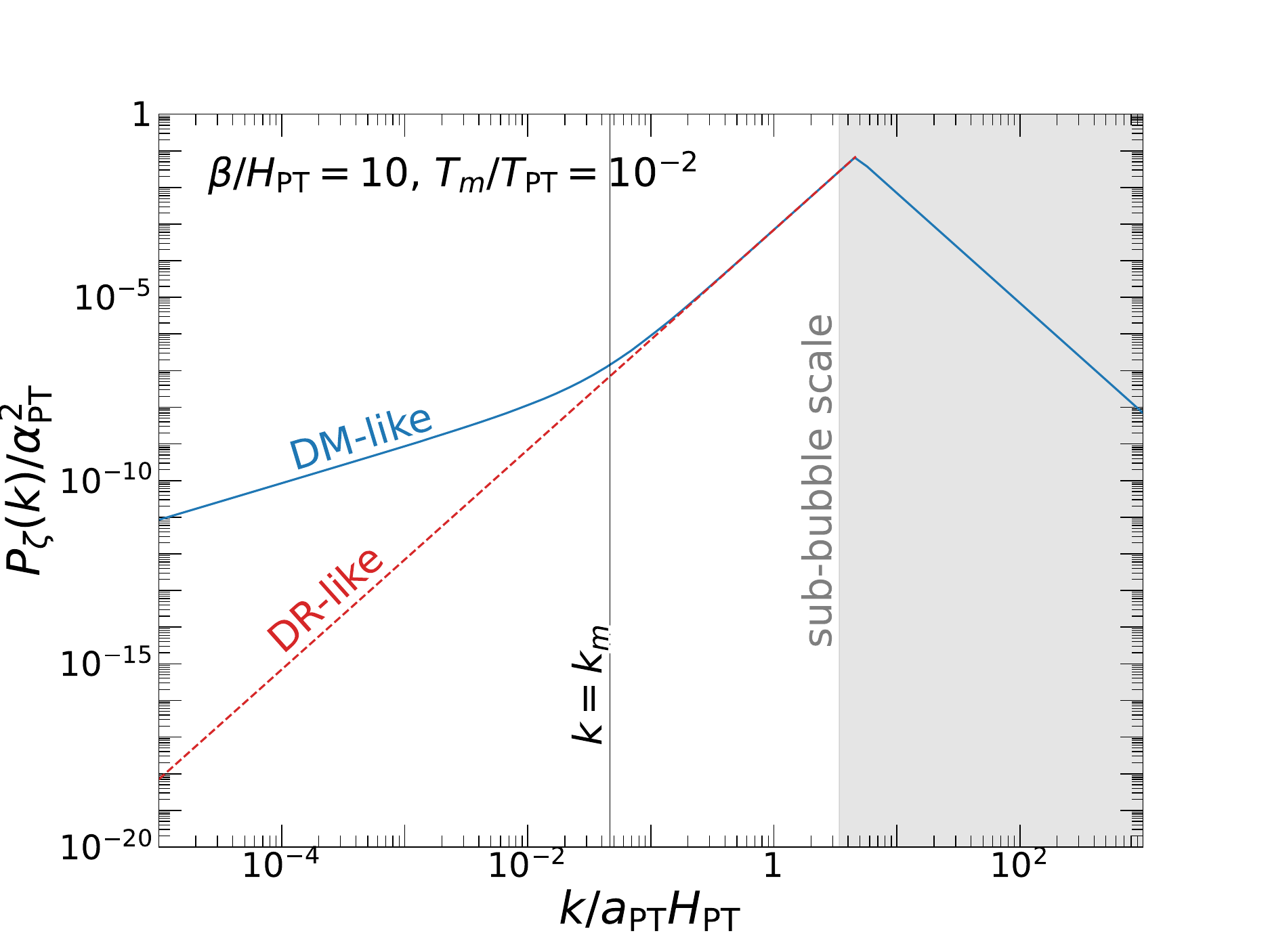}
    \caption{Curvature power spectrum from Eq.~\eqref{eq:Pzeta_full}, showing the enhancement at small $k$ for a DM-like dark sector, as opposed to the $k^3$ behavior one would find for dark radiation. The gray region corresponding to $k/a_{\rm PT} > \beta$ is subject to turbulence and hydrodynamical effects.}
    \label{fig:powerspectrum}
\end{figure}

Notice that for $t > t_m$, the DS energy density $\rho_d$ redshifts as $a^{-3}$, while $\rho_{\rm SM}$ redshifts as $a^{-4}$ during the radiation-dominated era. As a result, the relative contribution of the DS energy density grows as $\rho_d / \rho_{\rm SM} \propto \sqrt{t}$, leading to a corresponding enhancement of the curvature perturbation before horizon re-entry. Since the conformal time scales as $\tau/\tau_m = \sqrt{t/t_m}$ during radiation domination, the power spectrum acquires a $k^{-2}$ enhancement for small-$k$ modes that reenter the horizon after $t_m$. We find
\begin{equation}\label{eq:powerspectrumfinal}
    \mathcal{P}^{\rm DS}_\zeta(k) = \alpha_{\rm PT}^2 \mathcal{P}_{\delta t}(k) F(k_m/k)^2\,, 
\end{equation}
where $k_m = a_m H_m$ is the mode which enters the horizon at $t = t_m$. An example is shown in Fig.~\ref{fig:powerspectrum}, where the enhancement of the power spectrum at small $k$ is clearly visible. Since the PT-induced perturbations originate primarily from stochastic fluctuations in the PT completion time, they are uncorrelated with the primordial adiabatic perturbations. The total primordial power spectrum is therefore given by
\begin{equation}\label{eq:Ptotal}
\mathcal{P}^{\rm total}_\zeta(k)=\mathcal{P}^{\rm DS}_\zeta(k)+\mathcal{P}^{\rm AD}_\zeta(k)\,,
\end{equation}
where
\begin{equation}
\mathcal{P}^{\rm AD}_\zeta(k)=A_s\left(\frac{k}{k_*}\right)^{n_s-1}
\end{equation}
is the standard adiabatic contribution with $A_s = 2.101 \times 10^{-9}$, $k_* = 0.05 {\rm ~Mpc}^{-1}$, and $n_s = 0.9961$~\cite{Planck:2018vyg}.

%%%%%%%%%%%%%%%%%%%%%%%%%%%%%%%%%%%%%%%%%%%%%%%%%%%%%%%
%%%%%%%%%%%%%%%%%%%%%%%%%%%%%%%%%%%%%%%%%%%%%%%%%%%%%%%
%%%%%%%%%%%%%%%%%%%%%%%%%%%%%%%%%%%%%%%%%%%%%%%%%%%%%%%
%%%%%%%%%%%%%%%%%%%%%%%%%%%%%%%%%%%%%%%%%%%%%%%%%%%%%%%

A few comments are in order. First, in the absence of the $1/k^2$ enhancement encoded by $F(k_m/k)$, one recovers the power spectrum for DR derived in~\cite{Elor:2023xbz}. This limit is relevant for the DM+DR scenario.

Second, as mentioned above, the sharp transition between relativistic and nonrelativistic behavior in Eq.~\eqref{eq:curvature} is unphysical. In Appendix~\ref{app:transition}, we derive the perturbation more carefully by consistently accounting for the redshifting of the DS energy density, which smooths out this transition. The final result is
\begin{equation}\label{eq:Pzeta_full}
\mathcal{P}^{\rm DS}_\zeta(k) = \alpha_{\rm PT}^2 \mathcal{P}_{\delta t}(k) \left[ \frac{3}{4} \left(1 + w(k) \right) \right]^2 F(k)^2, \quad F(k) = \frac{k_c}{k} \frac{\chi(k)}{\chi(k_c)} ,
\end{equation}
where $\chi$ denotes the ratio of the average energy of a DM particle to its mass, $w$ is the DM equation-of-state parameter, and $k_c$ is the mode that enters the horizon at $t=t_c$.

Third, in deriving Eq.~\eqref{eq:powerspectrumfinal}, we assumed $\rho_d \ll \rho_{\rm SM}$ during radiation domination. This approximation breaks down after matter-radiation equality, when the DS contribution is no longer negligible. Consequently, the power spectrum returns to a $k^3$ scaling for $k < k_{\rm eq} \sim 10^{-2}~{\rm Mpc}^{-1}$.

Finally, we confirm this result, originally derived within the separate-universe formalism, by solving the superhorizon evolution equation for the curvature perturbation sourced by the DM isocurvature perturbation. We present this derivation in Appendix~\ref{app:superhorizon}.

\subsection{Subsequent evolution of composite WDM perturbations}\label{sec:evolution}
%Subsequent evolution}
The evolution of the DM perturbations after they enter the horizon is characterized by a transfer function.
The main way in which our DM-only scenario differs from standard, cold DM is that free-streaming can wash out small-scale structure.
These effects can be mapped onto those of thermal relic WDM, analogous to sterile neutrino scenarios~\cite{Bode:2000gq,Viel:2005qj}. To see this, recall that the relic density of a thermal warm relic with mass $m_{\rm th}$ and temperature $T_{\rm th}$ is
\begin{equation}\label{eq:thermaldensity}
    \Omega_{\rm th} h^2 = \left( \frac{T_{\rm th}}{T_\nu} \right)^3 \frac{m_{\rm th}}{94 {\rm ~eV} }\,,
\end{equation}
where $T_\nu = (4/11)^{1/3} T$ is the SM neutrino temperature. Free-streaming effects on the power spectrum depend only on $\Omega_{\rm th}h^2$ and $m_{\rm th} T_\nu / T_{\rm th}$. In our model, the corresponding quantity is
\begin{equation}
    \left( \frac{4}{11} \right)^{1/3} m_{\rm DM} \frac{T_{\rm PT}}{T_d(T_{\rm PT})} = \left( \frac{4}{11} \right)^{1/3} T_m \left[ \frac{g_*(T_{\rm PT})}{g_*(T_m)} \right]^{1/4} \approx \left( \frac{4}{11} \right)^{1/3} T_m\,,
\end{equation}
where we used Eq.~\eqref{eq:tm}. Equating this expression to $m_{\rm th} T_\nu / T_{\rm th}$ and setting $\Omega_{\rm th} = \Omega_{\rm DM}$ leads to 
\begin{equation}\label{eq:mapping}
    T_m = \left( \frac{11}{4} \right)^{1/3} m_{\rm th} \frac{T_\nu}{T_{\rm th}} = 6.2 {\rm ~keV} \left( \frac{m_{\rm th}}{1 {\rm ~keV}} \right)^{4/3} \left( \frac{\Omega_{\rm DM} h^2}{0.12} \right)^{-1/3} \,.
\end{equation}
 In the second equality, we used Eq.~\eqref{eq:thermaldensity} to solve for $T_\nu/T_{\rm th}$. The transfer function in our model for a given $T_m$ is identical to that of thermal WDM with mass $m_{\rm th}$ given by Eq.~\eqref{eq:mapping}, and consequently exhibits the same free-streaming suppression of the matter power spectrum.

\subsection{Compensation of warm DM suppression by PT perturbations}\label{sec:numerics}
When studying the effect of the PT on cosmological observables, we use the corrected curvature power spectrum $\mathcal{P}^{\rm DS}_\zeta (k)$ from Eq.~\eqref{eq:Pzeta_full} with the approximate peak shape $\mathcal{P}_{\delta t}$ given by Eq.~\eqref{eq:Pdt}. The curvature spectrum generated by the PT can be specified with four parameters: 
\begin{itemize}
    \item {$z_{\rm PT}$}: Redshift of the PT, directly related to the SM photon temperature $T_{\rm PT}$. For $T_{\rm PT} \lesssim 100 {\rm ~keV}$, we use the simple conversion formula $T_{\rm PT} = T_0 (1 + z_{\rm PT})$ 
    where $T_0 = 2.349\times 10^{-4}\ {\rm eV}$ is the SM photon temperature today. 
    \item {$\alpha_{\rm PT}$}: Ratio of the energy density released in the PT to that of the SM radiation energy density at the PT.
    \item {$\beta/H_{\rm PT}$}: Phase transition rate normalized to the Hubble expansion rate at the PT.
    \item {$r = T_m/T_{\rm PT}$}: Ratio of $T_m$ to $T_{\rm PT}$, where $T_m$ is the SM photon temperature when the DM composites become non-relativistic.
\end{itemize}
To study the effect of this PT signal on the SM photon and matter perturbations, we use the Boltzmann solver \CLASS~\cite{lesgourgues:2011class,Diego:2011class}. We implement the PT-enhanced curvature spectrum $\mathcal{P}^{\rm DS}_\zeta(k)+\mathcal{P}^{\rm AD}_\zeta(k)$ as the primordial spectrum through the built-in \texttt{external\_Pk} module. In total we use seven parameters in \texttt{external\_Pk}: three $\Lambda$CDM parameters $(A_s, n_s, k_*)$ for $\mathcal{P}^{\rm AD}_\zeta(k)$, and the four PT parameters for $\mathcal{P}^{\rm DS}_{\zeta}(k)$ described above.
%When studying the CMB+BAO bound, we further allow the $\Lambda$CDM parameters to vary.

We implement the warm dark matter (WDM) effects of our model using the built-in \CLASS\ \texttt{ncdm} (non-cold DM) module with number of species $\tt N_{\rm ncdm} = 1$\footnote{Our treatment of WDM suppression differs from that of the white noise fluctuations studied in Ref.~\cite{Amin:2025dtd}, where the causal ($k^3$) peak is not erased by free streaming because randomly distributed finite-size dark matter waves remain randomly distributed after streaming. In our case, particles can free-stream out of initially overdense bubbles and further smooth the density field, so the conventional WDM free-streaming suppression is expected to apply. Quantifying possible deviations from this approximation requires a dedicated analysis and is left for future work.}.
In particular, the ${\tt w}_{\tt ncdm}$ parameter for the WDM relic abundance is controlled by the PT parameters according to Eq.~\eqref{eq:relicabundance}.
We fix the total DM relic abundance ${\tt w}_{\tt tot} = {\tt w}_{\tt cdm} + {\tt w}_{\tt ncdm} = 0.12$ as we vary the PT parameters. This results in the following cases: 
\\
\\
(i) When Eq.~\eqref{eq:relicabundance} gives $\Omega_{\rm DM} h^2 < 0.12$, we set ${\tt w}_{\tt ncdm}$ to that value. The DM produced by the PT is interpreted as a subcomponent of the total DM, with the remaining relic abundance being made up of the usual CDM. 
\\\\
(ii) When Eq.~\eqref{eq:relicabundance} gives $\Omega_{\rm DM} h^2 = 0.12$, all of the DM is produced by the PT. We set ${\tt w}_{\tt ncdm} = {\tt w}_{\tt tot}$ and ${\tt w}_{\tt cdm} = 0$. This occurs when the PT parameters satisfy   \begin{equation}\label{eq:minimal}
     \alpha_{\rm PT} = 0.67 \left(\frac{1 {~\rm eV}}{r\, T_{\rm PT}}\right)\,.
     \end{equation}
(iii) When Eq.~\eqref{eq:relicabundance} gives $\Omega_{\rm DM} h^2 > 0.12$, we set ${\tt w}_{\tt ncdm}={\tt w}_{\tt tot}$. To avoid overproducing DM, we assume that either the excess latent heat does not contribute to the DM density (e.g. it becomes DR), or the DM density is depleted by DS dynamics.
\\

We map the PT parameters to the \texttt{ncdm} parameter ${\tt T_{\rm ncdm}}$, which corresponds to the temperature ratio of the equivalent thermal relic to the SM photons. From Eq.~\eqref{eq:thermaldensity} and the first equality of Eq.~\eqref{eq:mapping}, it follows that
\begin{equation}\label{eq:Tncdm}
    {\tt T_{\rm ncdm}} = 0.25\left[\left(\frac{{\tt w}_{\tt ncdm}}{0.12}\right)\left(\frac{1}{r}\right)\left(\frac{1 {~\rm keV}}{T_{\rm PT}}\right) \right]^{1/4} .
\end{equation}
From the relic abundance ${\tt w}_{\tt ncdm}$ and temperature $\tt T_{\rm ncdm}$, \CLASS\ determines the thermal relic mass $\tt m_{\rm ncdm}$ (see Eq.~\eqref{eq:thermaldensity}).
We summarize the $\Lambda$CDM parameters and DM abundances in our \CLASS\ implementation in Table~\ref{tab:lcdm}.
 
\begin{table}
    \centering
    \begin{tabular}{|>{\centering\arraybackslash}p{3.2cm}|>{\centering\arraybackslash}p{3.2cm}|}\hline
       \CLASS\ Parameters  & $\Lambda$CDM Value                   \\\hline
       $\omega_{\rm b}$                 &$0.02238$             \\\hline
       $h$                              &$0.6781$                \\\hline
       $\tau_{\rm reio}$                &$0.05431$             \\\hline
       $A_s$                            &$2.101\times 10^{-9}$ \\\hline
       $n_s$                            &$0.9661$              \\\hline
       $k_*~[{\rm Mpc}^{-1}]$ &$0.05$                             \\\hline
    \end{tabular}
    
    \vspace{0.3 cm}
    
    \begin{tabular}{|>{\centering\arraybackslash}p{3.2cm}|>{\centering\arraybackslash}p{3.2cm}|>{\centering\arraybackslash}p{3.2cm}|>{\centering\arraybackslash}p{3.2cm}|}\hline
      \CLASS\ Parameters (DM Abundances) & $\Lambda$CDM Value (${\tt w}_{\tt tot} \equiv 0.12011$) &  (i) Eq.~\eqref{eq:relicabundance} $< {\tt w}_{\tt tot}$ & (ii, iii) Eq.~\eqref{eq:relicabundance} $ \geq {\tt w}_{\tt tot} $\\\hline
       ${\tt w}_{\tt ncdm}$ & $0$ &Eq.~\eqref{eq:relicabundance}        & ${\tt w}_{\tt tot}$\\\hline
       ${\tt w}_{\tt cdm}$ & ${\tt w}_{\tt tot}$ &${\tt w}_{\tt tot} - {\tt w}_{\tt ncdm}$ & $0$\\\hline
    \end{tabular}
    \caption{Settings used for $\Lambda$CDM parameters in \CLASS\ calculations. When non-cold DM is introduced, the $\Lambda$CDM value ${\tt w}_{\tt tot}$ for the CDM abundance is kept as the total DM abundance such that ${\tt w}_{\tt cdm} + {\tt w}_{\tt ncdm} = {\tt w}_{\tt tot}$ is held constant. In case (i) all latent heat is converted to WDM, which only makes up a fraction of the total DM with the remainder being the usual CDM. In cases (ii) and (iii) the WDM makes up all the DM with excess latent heat (if any) tacitly converted into something else.}
    \label{tab:lcdm}
\end{table}

By calculating the ratio of the predicted matter power spectrum to the $\Lambda$CDM result, we obtain the free-streaming suppression described at the end of Section~\ref{sec:evolution}. We will use this to correct the curvature perturbations when studying direct constraints on $\mathcal{P}^{\rm DS}_\zeta(k)$.  

\begin{figure}
    \centering
    \includegraphics[width=0.8\linewidth]{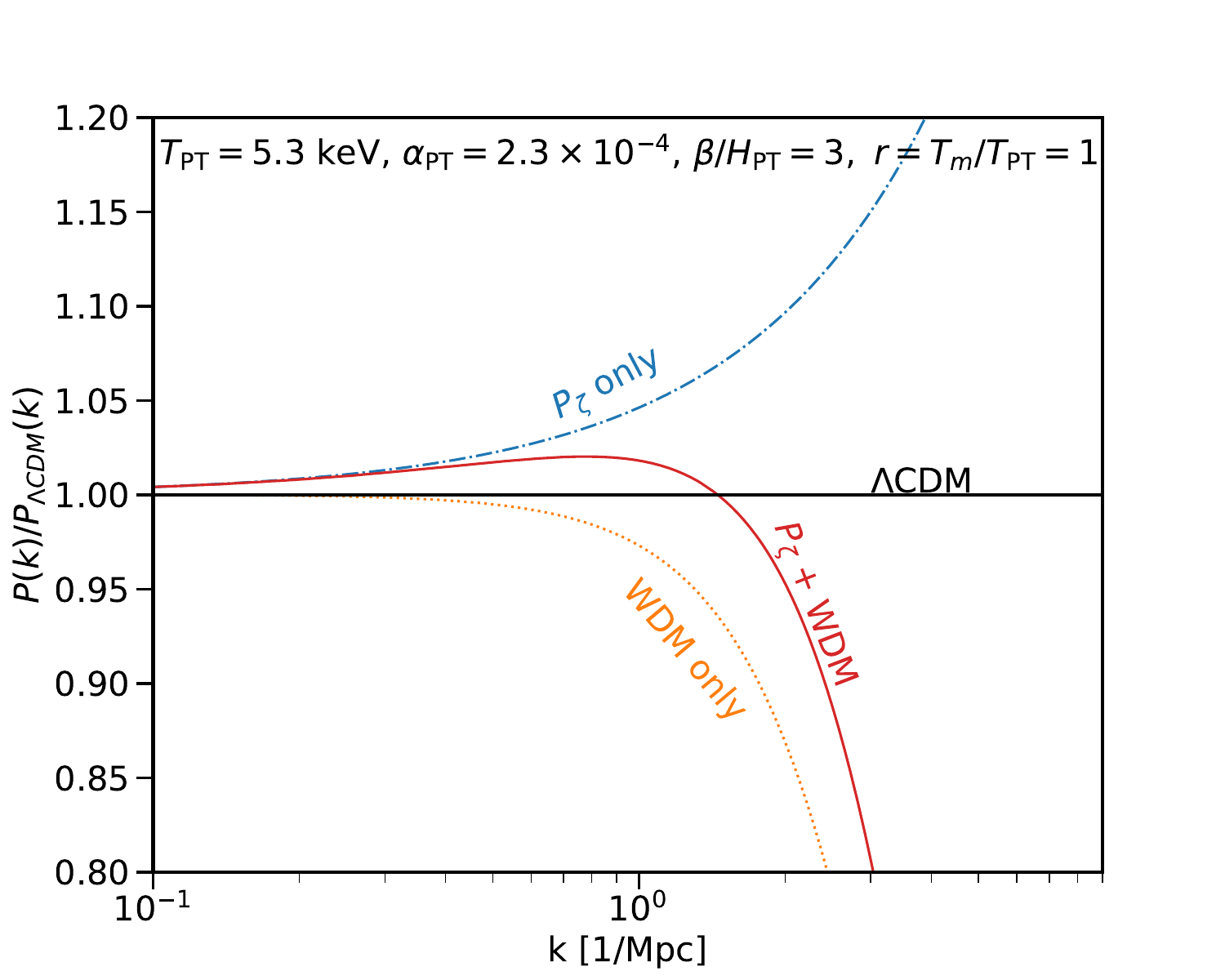}
    \caption{
    Ratio of linear matter power spectra from \CLASS\ with respect to $\Lambda$CDM at redshift $z=0$, for $T_{\rm PT} = 5.3{~\rm keV}$, $\alpha_{\rm PT} = 2.3\times 10^{-4}$, $\beta/H_{\rm PT} = 3$ and $r = 1$. The plot includes cases isolating the $P_{\zeta}$ peak enhancement (blue dot-dashed), the WDM free-streaming suppression (orange dotted), and a combination of both effects (solid red) to the matter power spectrum. The ratio is shown in linear scale, and we zoom in to emphasize the competition between the $\mathcal{P}_{\zeta}$ peak enhancement and the WDM free-streaming suppression for modes $k\sim 1\, h\,\mathrm {Mpc}^{-1}$.}
    \label{fig:matter_Pk}
\end{figure}

In Fig.~\ref{fig:matter_Pk}, we present an example of DS-induced corrections to the matter power spectrum. The ratio of the matter power spectrum to the $\Lambda$CDM prediction is shown as the solid red curve. In principle, one should include the full DM transfer function from \CLASS\ to correct for the growth of PT-induced modes that were sub-horizon before the PT (see Ref.~\cite{Greene:2026gnw} for details). However, since the modes we focus on in our \CLASS\ analysis are superhorizon at the time of the PT, these corrections are expected to be mild.

To isolate different effects on the matter power spectrum, we also plot a scenario including the $\mathcal{P}^{\rm DS}_{\zeta}(k)$ enhancement of the curvature perturbation but neglecting WDM effects (blue dot-dashed line), as well as a scenario including only the WDM free-streaming suppression without the PT-enhanced curvature perturbation (orange dotted line). These two effects compete in our model: as $k$ increases, the matter power spectrum grows as we approach the PT scale due to the $\propto k$ infrared tail of $\mathcal{P}^{\rm DS}_{\zeta}(k)$, until free-streaming takes over to suppress the power spectrum at higher $k$.

In particular, the example in Fig.~\ref{fig:matter_Pk} is for a DS scenario in which the matter power spectrum (solid red) satisfies the compressed eBOSS Lyman-$\alpha$ constraint at the pivot scale $k \approx 1\,h\,\mathrm{Mpc}^{-1}$ (see Sec.~\ref{sec:Lymana} for details), while the WDM-only case (orange), which does not include the enhancement from $\mathcal{P}^{\rm DS}_\zeta$, is excluded. The data are sensitive to the shape of the matter power spectrum; the enhancement in $\mathcal{P}^{\rm DS}_\zeta$ partially compensates for the WDM suppression, preserving a shape closer to the $\Lambda$CDM spectrum up to higher $k$-modes and thereby relaxing the WDM constraint\footnote{A similar compensation of the WDM suppression in the matter power spectrum is studied in~\cite{Tadepalli:2025gzf,Co:2025hbi}, where CDM isocurvature perturbations relax the WDM constraint. In contrast, we consider enhanced curvature perturbations that modify both CDM and WDM fluctuations, with a concrete realization from a DS confinement PT.} .

%%%%%%%%%%%%%%%%%%%%%%%%%%%%%%%%%%%%
%%%%%%%%%%%%%%%%%%%%%%%%%%%%%%%%%%%
\section{Analysis}\label{sec:analysis}
We analyze CMB and structure formation constraints on the PT-induced perturbations to derive $2\sigma$ exclusion bounds on the energy density ratio $\alpha_{\rm PT}$. Although the parameter scans are more conveniently performed in terms of redshift $z_{\rm PT}$, we present the bounds in terms of the temperature $T_{\rm PT}$, which is more physically intuitive. 
We derive constraints in the $(T_{\rm PT}, \alpha_{\rm PT})$ plane for fixed values of $\beta/H_{\rm PT}$ and $r=T_m/T_{\rm PT}$. In this parameter space, the DM-only scenario with the total DM abundance comprised of the composite DM corresponds to a curve determined by Eq.~\eqref{eq:minimal}.

To obtain the DS scenario with maximal curvature perturbations, we consider a supercooled first-order PT with $\beta/H_{\rm PT}=3$. Tensor perturbations from FOPTs with similarly small $\beta/H_{\rm PT}$ have been studied in~\cite{Zhong:2021hgo,Yamada:2025hfs,Lewicki:2025hxg}, where redshifting effects modify the power spectrum amplitude and peak location by $\mathcal{O}(1)$ factors. We therefore expect our bounds on $\alpha_{\rm PT}$ and $T_{\rm PT}$ to be rescaled by comparable amounts. This effect does not have a qualitative impact on our results, so we leave a detailed study of it to future work. 
We further consider $r=1$, i.e. the composite DM produced in the PT becomes nonrelativistic immediately after the transition.

For the DM-only scenario, we concentrate on the range $T_{\rm PT}\in(1,10)$~keV, where perturbation signals from the composite WDM sourced by the PT are most significant. In this region, we perform detailed likelihood analyses using CMB anisotropy and Lyman-$\alpha$ data.
The assumption that the DS is dominated by DM leads to the strongest possible constraints through the matter enhancement of $\mathcal{P}_\zeta^{\rm DS}$. We remark that this is challenging to realize for values of $\alpha_{\rm PT}$ above the relic abundance line, for which one would generically overproduce DM. One could imagine, for instance, that a subcomponent of the DM decays to DR immediately after the $k$-modes probed by Lyman-$\alpha$ data enter the horizon. The point is that one should be careful in interpreting the bounds we derive above the relic abundance line.

A more plausible scenario for larger values of $T_{\rm PT}$ and $\alpha_{\rm PT}$ is the DM+DR scenario, in which the excess latent heat not converted into DM is transferred into DR. In this case, the dominant correction to curvature perturbation arises from the DR, and the resulting constraints are expected to closely resemble those obtained for DR scenarios in Ref.~\cite{Elor:2023xbz, Greene:2026gnw}.

\subsection{CMB+BAO}
We conduct a likelihood study of the CMB anisotropy constraints for $T_{\rm PT} \in (1, 10)$~keV using a Markov Chain Monte Carlo (MCMC) analysis with {\it Planck} and BAO datasets. Since the sensitivity of the {\it Planck} measurement is limited to perturbation modes $k\lesssim 0.1{~\rm Mpc}$, there are some approximations we can make to speed up the \CLASS\ theory code calculation.
First, {\it Planck} is sensitive only to modes that remain superhorizon during the PT considered here, allowing us to treat the PT-induced perturbations as initial conditions when computing the CMB signals. 
Second, WDM effects are negligible on these scales: for $T_{\rm PT} \gtrsim 520\,\mathrm{eV}$, the corrections to the $C_\ell^{\rm TT,TE,EE}$ power spectra are below the percent level for the $\ell \leq 2500$ modes probed by {\it Planck}. We therefore neglect WDM effects in this MCMC analysis, although they are included later in the Lyman-$\alpha$ forest analysis. Using the example illustrated in Fig.~\ref{fig:matter_Pk} as a guide, we study the CMB+BAO constraint with the ``$P_\zeta$ Only'' scenario as a good approximation to the full result.

The parameter space of the scan consists of six $\Lambda$CDM parameters and two additional parameters $z_{\rm PT}$ and $\alpha_{\rm PT}$, which we scan over in log space. The parameters and their prior ranges are given in Table \ref{tab:prior}.  
\begin{table}
    \centering
    \begin{tabular}{|c|c|}\hline
       $\Lambda$CDM Parameters  & Range  \\\hline
       $10^2 \omega_{\rm b}$    &$[1.8, 3]$      \\\hline
       $\omega_{\rm cdm}$       &$[0.1, 0.2]$    \\\hline
       $H_0 \rm[km/s/Mpc]$      &$[60.0, 80.0]$  \\\hline
       $\tau_{\rm reio}$        &$[0.004, 0.12]$ \\\hline
       $10^9A_s$                &$[1.8, 3]$      \\\hline
       $n_s$                    &$[0.9, 1.1]$    \\\hline
    \end{tabular}
    
    \vspace{0.3 cm}
    
    \begin{tabular}{|c|c|}\hline
       PT Parameters  & Range ($\beta/H_{\rm PT}= 3, r=1$)\\\hline
       $\log_{10}(z_{\rm PT})$  &$[6, 8]$        \\\hline
       $\log_{10}(\alpha_{\rm PT})$  &$[-5, -3]$ \\\hline
    \end{tabular}
    \caption{Parameters and prior ranges used for MCMC scan with the six $\Lambda$CDM parameters allowed to vary. For the PT parameters, we fix $\beta/H_{\rm PT}= 3$ and $ r=1$ and scan over $z_{\rm PT}$ and $\alpha_ {\rm PT}$ in log space (base 10). Note that our approximation of neglecting WDM effects is only valid for $z_{\rm PT} \gtrsim 2.2\times 10^6$, so we have to truncate the lower $z_{\rm PT}$ region of the bound. This is automatically satisfied when we restrict to $T_{\rm PT}\geq 1{\rm~keV}$ in the final result.}
    \label{tab:prior}
\end{table}
We use a combination of the following datasets:
 \begin{itemize}
     \item \underline{Planck}: Measurements of the CMB temperature and polarization anisotropies from the {\it Planck} 2018 dataset, which includes the low-$\ell$ $(\ell < 30)$ TT, EE and high-$\ell$ $(\ell \geq 30)$ TTTEEE measurements~\cite{Planck:2019nip}. We also include the {\it Planck} Lensing likelihood~\cite{Planck:2018lbu}.
     \item \underline{BAO}: Baryon Acoustic Oscillation (BAO) measurements, which include the 6DF Galaxy survey, SDSS-DR7 MGS data, and the BOSS measurement of the BAO scale and $f\sigma 8$ from the DR12 galaxy sample~\cite{Beutler_2011, Ross:2014qpa, BOSS:2016wmc}.
 \end{itemize}
We perform the MCMC analysis using \texttt{MontePython}~\cite{Audren:2012wb,Brinckmann:2018cvx} with the Metropolis--Hastings algorithm~\cite{Hastings:1970aa}. Following~\cite{Planck:2018vyg}, we adopt the convention of modeling the free-streaming neutrinos as two massless and one massive species with mass $0.06$ eV. The Gelman--Rubin convergence criterion $R-1 < 0.0163$ is satisfied \cite{Rubin:1992}; we analyze and plot the MCMC samples using \texttt{GetDist}~\cite{Lewis:2019xzd}. 

\begin{figure}
    \centering
    \includegraphics[width=0.7\linewidth]{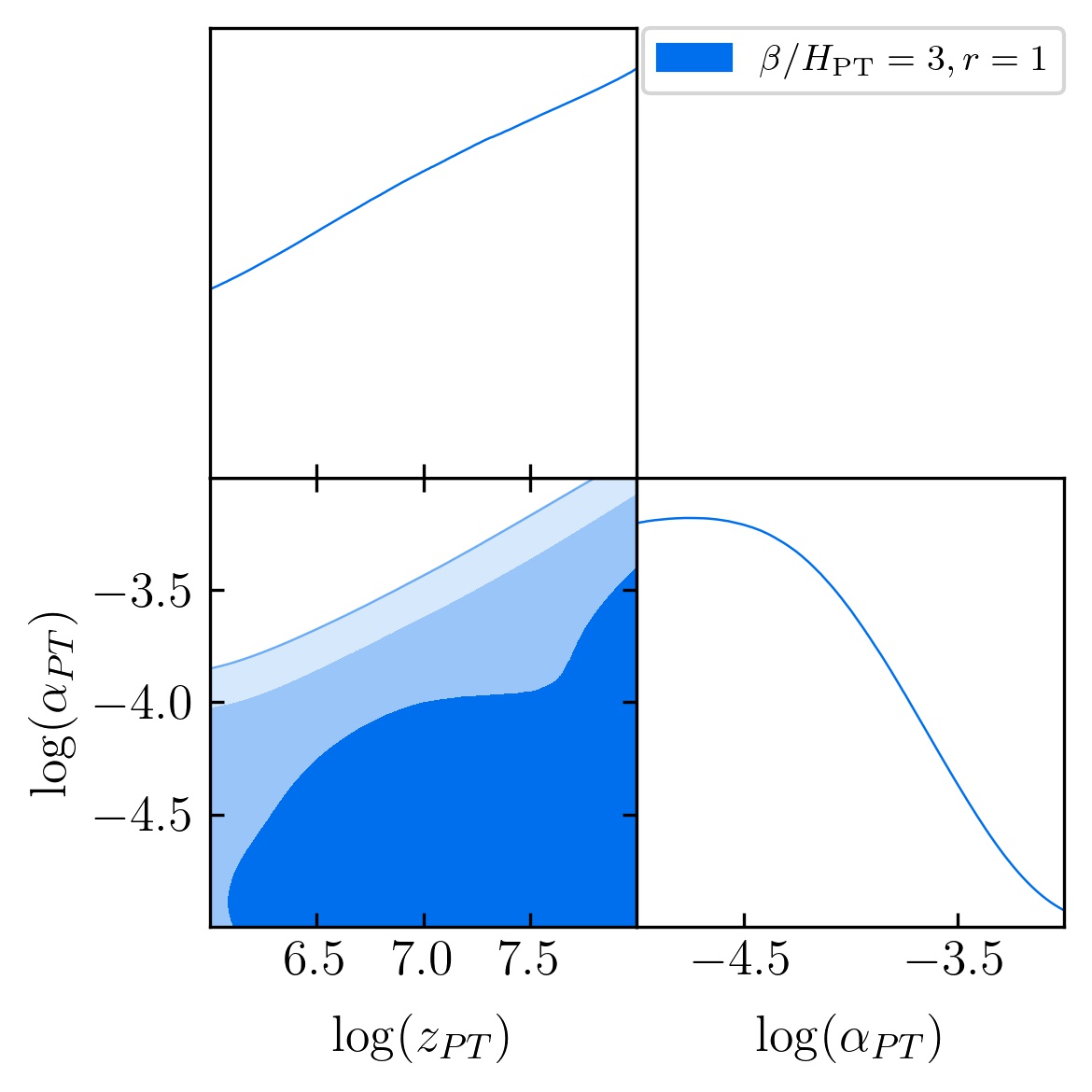}
    \caption{Results of the MCMC analysis with CMB+BAO data for $\beta/H_{\rm PT}= 3$ and $r= 1$, marginalized to $z_{\rm PT}$ and $\alpha_{\rm PT}$ in log space (base $10$). The 2D contour plots show the preferred regions up to 1, 2 and 3$\sigma$ boundaries. The MCMC analysis was performed while neglecting WDM free-streaming effects, which is a valid approximation for $z_{\rm PT} \geq 2\times 10^6$ when the WDM effects are beyond the {\it Planck} sensitivity. This condition is automatically satisfied for $T_{\rm PT}$ in the keV scale.}
    \label{fig:MCMCtriangle}
\end{figure}

Fig.~\ref{fig:MCMCtriangle} shows the preferred regions for $\beta/H_{\rm PT}= 3$ and $r = 1$, marginalized to $z_{\rm PT}$ and $\alpha_{\rm PT}$. 
%The 2D contours in the $(z_{\rm PT}\ ,\alpha_{\rm PT})$ plane are plotted up to $3 \sigma$.
Note that the MCMC scan runs over an extended range of redshifts $10^6 \leq z_{\rm PT} \leq 10^8$, which translates approximately $235{~\rm eV} \lesssim T_{\rm PT} \lesssim 23.5{~\rm keV}$. Hence, there is a region $z_{\rm PT} \lesssim 2.2\times 10^6$ where the approximation of neglecting WDM effects breaks down, so one should exercise caution when interpreting the figure in this region. Nonetheless, this point has no bearing on our main results in Section~\ref{sec:results}, as we truncate to  $T_{\rm PT} \in (1, 10)$~keV. 

\subsection{Lyman-$\alpha$}\label{sec:Lymana}
We estimate constraints for $T_{\rm PT} \in (1, 10)$~keV by conducting a reduced Lyman-$\alpha$ likelihood study. For each point in the PT parameter space, this involves calculating the corresponding linear matter power spectrum from \texttt{CLASS} and fitting it against the compressed eBOSS data at the Lyman-$\alpha$ pivot scale. The compressed data consist of two parameters~\cite{McDonald:2000,Chabanier:2019,Pedersen:2020,Pedersen:2021,Pedersen:2023,Goldstein:2023}:   
\begin{equation}
\Delta^2_{\rm lin} \equiv \frac{k^3_p}{2\pi^2} {\rm P}_{\rm lin}(k_p,z_p), \qquad\qquad n_{\rm lin} \equiv \left.\frac{{\rm d}\ln {\rm P}_{\rm lin}(k,z)}{{\rm d}\ln k}\right|_{k_p,z_p},
\end{equation}
where $\Delta^2_{\rm lin}$ denotes the amplitude and $n_{\rm lin}$ the logarithmic tilt of the linear matter power spectrum ${\rm P}_{\rm lin}(k,z)$, evaluated at redshift $z_p = 3$ and pivot wavenumber $0.009\ {\rm s/km}$ (in velocity units). This pivot wavenumber translates to the pivot scale $k_p \approx 1.03{h\rm~Mpc}^{-1}$ in conformal units~\cite{Bagherian:2024,Bansal:2024,Buckley:2025}. We use the following data for our reduced 2-parameter likelihood study: 
\begin{itemize}
    \item \underline{PRIYA}: A compressed likelihood using Lyman-$\alpha$ measurements of the 1D flux power spectrum from SDSS DR14 BOSS and eBOSS quasars~\cite{Bird:2023,Fernandez:2024,He:2025}. Following Ref.~\cite{Bird_github:2026}, we construct a 2D Gaussian likelihood using the data points $\Delta^2_{\rm lin} = 0.267 \pm 0.022$ and $n_{\rm lin} = -2.288 \pm 0.024$ with correlation coefficient $0.4$ between them.
 \end{itemize}

We conduct the reduced likelihood study using the \CLASS\ setup described in Section~\ref{sec:numerics}, including the implementation of WDM effects in the \texttt{ncdm} module. Three comments are in order.
First, the pivot scale $k_p$ is always superhorizon at the PT time, so we can neglect transfer function corrections for any evolution before the PT.
Second, since the reduced likelihood requires the derivative of the matter power spectrum ${\rm P}_{\rm lin}(k,z)$ to be well-defined at $k_p$, we have to make sure that there are no sharp oscillations in ${\rm P}_{\rm lin}(k,z)$ at $k_p$, which can occur with WDM effects. Fortunately, such features only appear at $k > k_p$ for a keV-scale PT, as illustrated in Fig~\ref{fig:matter_Pk}.
Finally, \texttt{ncdm} calculations in this PT parameter range require optimizing \CLASS\ precision settings to avoid numerical artifacts in ${\rm P}_{\rm lin}(k,z)$ that interfere with the fit to the data\footnote{To ensure accurate numerical integration and sampling, we adopt the precision settings specified in the \texttt{pk\_ref.pre} file of \CLASS\,, refining both the Einstein--Boltzmann solver and the sampling step size.}. The optimization increases the calculation time significantly, making a full MCMC scan unwieldy. Therefore, when deriving constraints on the PT parameters, we compute the $\chi^2$ directly (described below) with the $\Lambda$CDM parameters fixed, without performing a full MCMC scan.

\begin{table}
    \centering     
    \begin{tabular}{|c|c|c||c|c|}\hline
        Scenario&  DM Type&Curvature $\mathcal{P}(k)$ & $T_{\rm PT} {~\rm [keV]}$&$\alpha_{\rm PT}$\\\hline
        $\Lambda{\rm CDM}$& CDM&$\mathcal{P}_\zeta^{\rm AD}$& --&--\\\hline
        $P_\zeta$ Only& CDM&$\mathcal{P}_\zeta^{\rm AD} + \mathcal{P}_\zeta^{\rm DS}$&$[1, 10]$ &$[10^{-4}, 10^{-3}]$\\\hline
  WDM Only&  \texttt{ncdm}&$\mathcal{P}_\zeta^{\rm AD}$& $[1, 10]$ &$[10^{-5}, 5\times10^{-4}]$\\\hline
  $P_\zeta + {\rm WDM}$&  \texttt{ncdm}&$\mathcal{P}_\zeta^{\rm AD} + \mathcal{P}_\zeta^{\rm DS}$& $[1, 10]$ &$[10^{-5}, 10^{-3}]$\\\hline
    \end{tabular}
    \caption{Scenarios and parameter ranges considered for the reduced Lyman-$\alpha$ likelihood study of the linear matter power spectrum. The $\Lambda$CDM parameters are fixed according to the \CLASS\ settings from Table~\ref{tab:lcdm}. A DM type of ``CDM'' indicates that the CDM abundance was fixed to the $\Lambda$CDM value with WDM effects neglected, while ``\texttt{ncdm}'' indicates that WDM  effects were implemented in the \CLASS\ \texttt{ncdm} module following the convention in Table~\ref{tab:lcdm}. The Curvature $\mathcal{P}(k)$ column shows the contributions included in the primordial curvature power spectrum for each scenario. The PT parameters $T_{\rm PT}$ and $\alpha_{\rm PT}$ were looped over in log-spaced steps for fixed $\beta/H_{\rm PT} = 3$ and $r = 1$.}
    \label{tab:class}
\end{table}

We perform the \CLASS\ calculations fixing the $\Lambda$CDM parameters according to Table~\ref{tab:lcdm}. As in Section~\ref{sec:numerics}, we study scenarios which isolate the enhancement of the power spectrum by the PT and the suppresion by WDM effects, in addition to our full model which incorporates both effects. This allows us to better understand the form of the Lyman-$\alpha$ bound in different regions of parameter space. For each of these scenarios, we fix $\beta/H_{\rm PT} = 3$ and $r=1$  while varying $T_{\rm PT}$ and $\alpha_{\rm PT}$ on a log scale. The different scenarios and parameter ranges are listed in Table~\ref{tab:class}. 
For each PT parameter choice, we fit ${\rm P}_{\rm lin}(k, z_p)$ to the compressed data to compute $\chi^2_{\rm PT}$. We evaluate the goodness of fit by calculating the $\Delta \chi^2$ with respect to $\Lambda$CDM:
\begin{equation}
    \Delta\chi^2 = \chi_{\rm PT}^2 - \chi_{\Lambda\rm CDM}^2 .
\end{equation}
Since we perform the fit by iterating over $\alpha_{\rm PT}$ for each choice of $T_{\rm PT}$, we construct the $2\sigma$ region by keeping points within  $\Delta\chi^2 \leq 3.84$ (corresponding to $p = 0.05$ for one degree of freedom).

\begin{figure}
    \centering
    \includegraphics[width=0.8\linewidth]{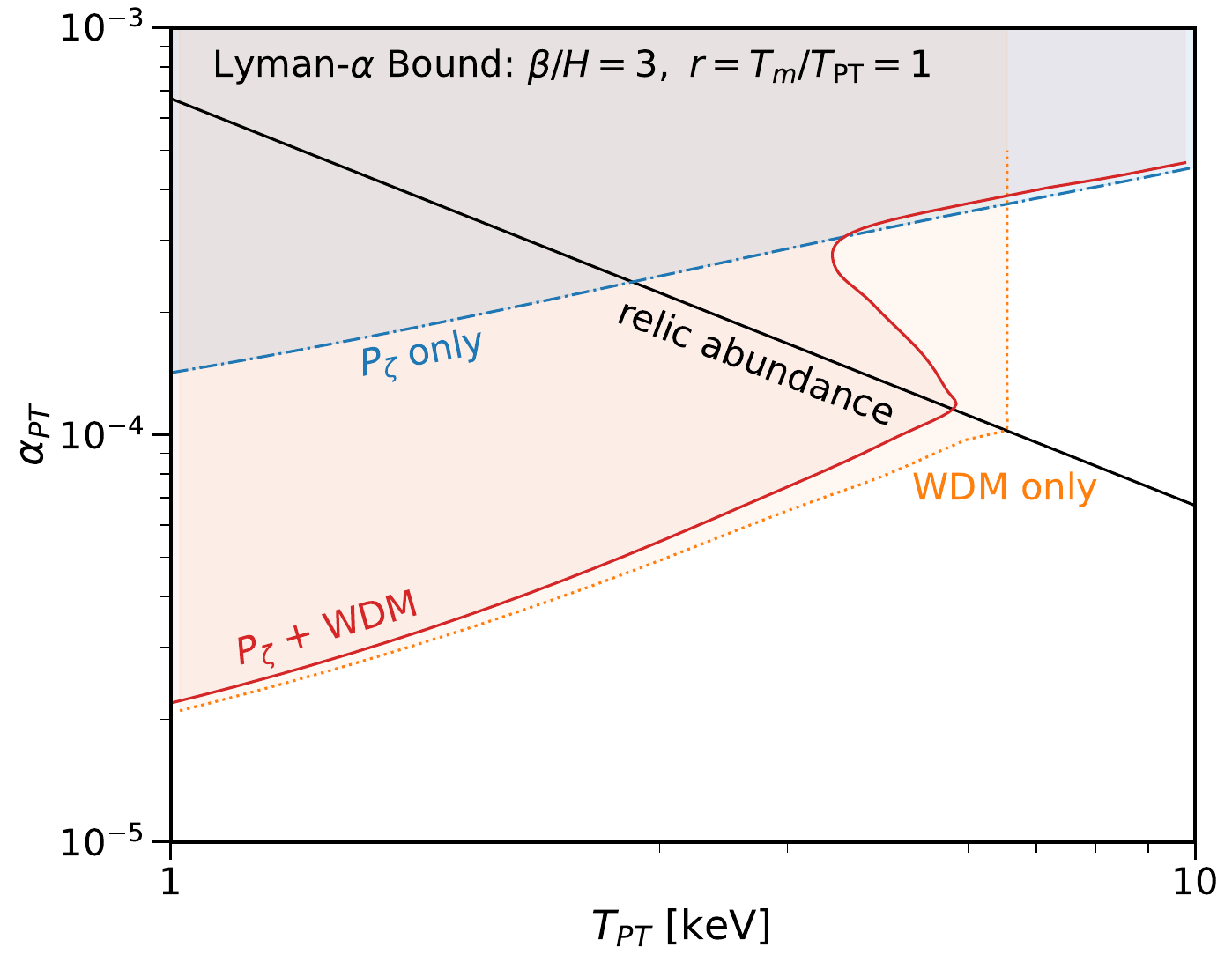}
\caption{$2\sigma$ exclusion bound from compressed Lyman-$\alpha$ likelihood, given in $T_{\rm PT}$ and $\alpha_{\rm PT}$ for fixed $\beta/H_{\rm PT}=3$ and $r=1$, assuming the DS energy density is dominated by composite DM. We consider the $\chi^2$ goodness of fit with respect to $\Lambda$CDM, studying constraints on the matter power spectrum for scenarios where we separately isolate the $\mathcal{P}_\zeta(k)$ peak enhancement (blue) and the WDM free-streaming suppression (orange), as well as our DS model  which includes both effects (red). The black line delineates the PT parameters that yield the observed relic abundance in the DM-only scenario (Eq.~\eqref{eq:minimal}). As described in Table~\ref{tab:lcdm}, the total DM abundance is held constant such that the WDM comprises all the DM above the line but is a DM subcomponent below the line.}
    \label{fig:lyman}
\end{figure}

Fig.~\ref{fig:lyman} shows the $2\sigma$ exclusion bounds in $T_{\rm PT}$ and $\alpha_{\rm PT}$. For reference, we also plot the minimal relic abundance line described by Eq~\eqref{eq:minimal}.
The orange line in Fig.~\ref{fig:lyman} is the bound considering WDM effects but not the enhancement of the power spectrum. Below the relic abundance line, the WDM constitutes a fraction of the total DM proportional to $\alpha_{\rm PT}$. Hence, as $\alpha_{\rm PT}$ decreases, the WDM effects are diluted and the lower bound on $T_{\rm PT}$ decreases. Above the relic abundance line, the WDM abundance is fixed, so that the WDM effects only depend on $r \,T_{\rm PT}$ (see Eq.~\eqref{eq:Tncdm}). Thus the bound is a line of constant $T_{\rm PT}$.
The blue line in Fig.~\ref{fig:lyman} is the bound on the enhanced matter power spectrum, originating from the infrared tail of $\mathcal{P}_\zeta^{\rm DS}(k)$, and ignoring WDM effects.

The red line in Fig.~\ref{fig:lyman}, which describes the Lyman-$\alpha$ exclusion bound for our minimal model, includes both effects.
Correspondingly, it interpolates between the bounds including only the effect of WDM or an enhanced power spectrum. It is closer to the former at lower $T_{\rm PT}$ where WDM effects dominate, but approaches the latter at higher $T_{\rm PT}$ as the enhanced curvature perturbation becomes more important.
At intermediate temperatures, $T_{\rm PT}\sim 4$~keV, a larger $\alpha_{\rm PT}$ enhances the power spectrum and can compensate for the WDM suppression around the pivot scale $k_p$, leading to the characteristic ``indentation" feature in the bound.

\subsection{Other cosmological constraints}\label{sec:otherbounds}
For DS phase transitions that predominantly reheat into DM, we focus on the CMB and Lyman-$\alpha$ constraints discussed above. For models in which the latent heat is instead transferred primarily into DR, like the DM+DR scenario  with $\alpha_{\rm PT}$ well above the relic-abundance curve, we also study the following constraints on the PT perturbations.

We consider only the DR contribution to $\mathcal{P}^{\rm DS}_\zeta$ and compute the power spectrum using Eq.~\eqref{eq:powerspectrumfinal}, omitting the $F(k_m/k)$ term that arises from the slower redshift of the composite DM\footnote{We again ignore the correction for the growth of modes that were sub-horizon before the PT~\cite{Greene:2026gnw}. The effect of this correction on our results is negligible for $\beta/H = 3$.}. We also neglect free-streaming suppression since we assume the DM is cold. We then confront the resulting $\mathcal{P}^{\rm total}_\zeta$ with existing bounds, including constraints from CMB spectral distortions~\cite{Chluba:2012we,Cyr:2023pgw}, dynamical heating of stars in ultra-faint dwarfs~\cite{Graham:2024hah}, early reionization~\cite{Qin:2025ymc}, formation of ultracompact minihalos~\cite{Bringmann:2025cht}, and pulsar timing arrays~\cite{Lee:2020wfn}. Requiring consistency with these limits yields an upper bound on $\alpha_{\rm PT}$. The resulting constraints are similar to those obtained in~\cite{Elor:2023xbz,Greene:2026gnw} for a DS phase transition that reheats into DR.

The dark PT also contributes to the effective number of neutrino species $N_{\rm eff}$.
Immediately after the phase transition, the dark sector energy density is
\begin{equation}
    \rho_d(T_{\rm PT}) = \alpha_{\rm PT} \frac{\pi^2}{30} g_*(T_{\rm PT}) T_{\rm PT}^4 = \frac{1}{2} \alpha_{\rm PT}\,g_*(T_{\rm PT}) \rho_\gamma(T_{\rm PT}) ,
\end{equation}
with $\rho_\gamma$ the photon energy density. Comparing this to the neutrino energy density $\rho_\nu$, we identify the change to $N_{\rm eff}$:
\begin{equation}
    \Delta N_{\rm eff} = \frac{4}{7} \left( \frac{11 g_{*,s}(T_{\rm CMB})}{4 g_{*,s}(T_{\rm PT})} \right )^{4/3} \alpha_{\rm PT}\, g_*(T_{\rm PT})\, .
\end{equation}
In our main results we will consider the current bound $\Delta N_{\rm eff} < 0.29$ \cite{Planck:2018vyg} and a projection for the Simons Observatory, $\Delta N_{\rm eff} < 0.1$ \cite{SO:2019}.

%%%%%%%%%%%%%%%%%%%%%%%%%%%%%%%%%%%%%%%%%%%%%%%%%%%%%%%
%%%%%%%%%%%%%%%%%%%%%%%%%%%%%%%%%%%%%%%%%%%%%%%%%%%%%%%
\section{Results}\label{sec:results}
%%%%%%%%%%%%%%%%%%%%%%%%%%%%%%%%%%%%%%%%%%%%%%%%%%%%%%%
%%%%%%%%%%%%%%%%%%%%%%%%%%%%%%%%%%%%%%%%%%%%%%%%%%%%%%%

\begin{figure}
    \centering
    \includegraphics[width=0.8\textwidth]{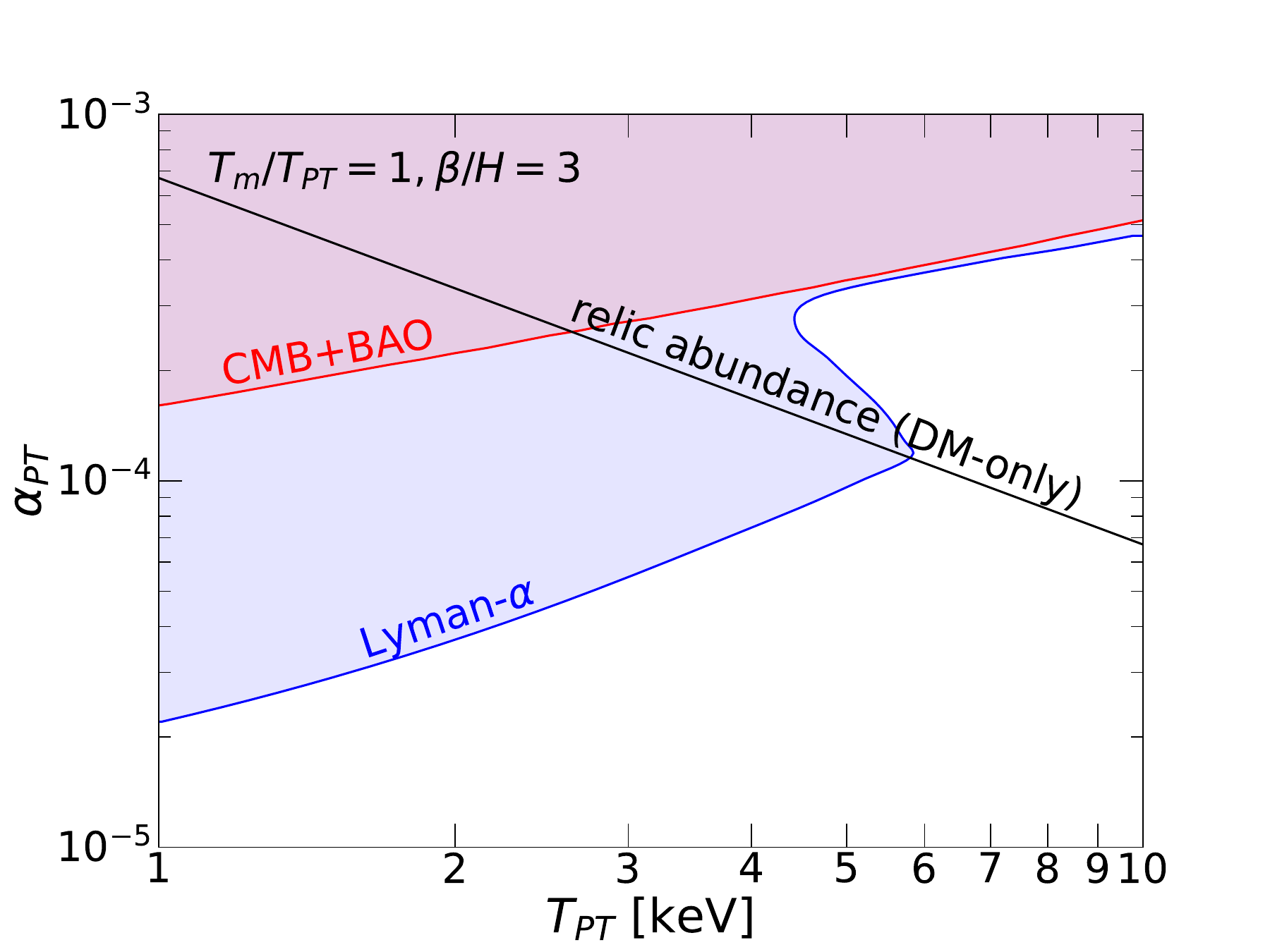}
    \caption{Bounds on the dark PT for an optimistic benchmark $r = 1, \beta/H_{\rm PT} = 3$. We zoom in on the $1$--$10$~keV region and assume the DM abundance is set by the DM-only scenario described in the main text.}
    \label{fig:exclusion_zoom}
\end{figure}

\begin{figure}
    \centering
    \includegraphics[width=0.8\textwidth]{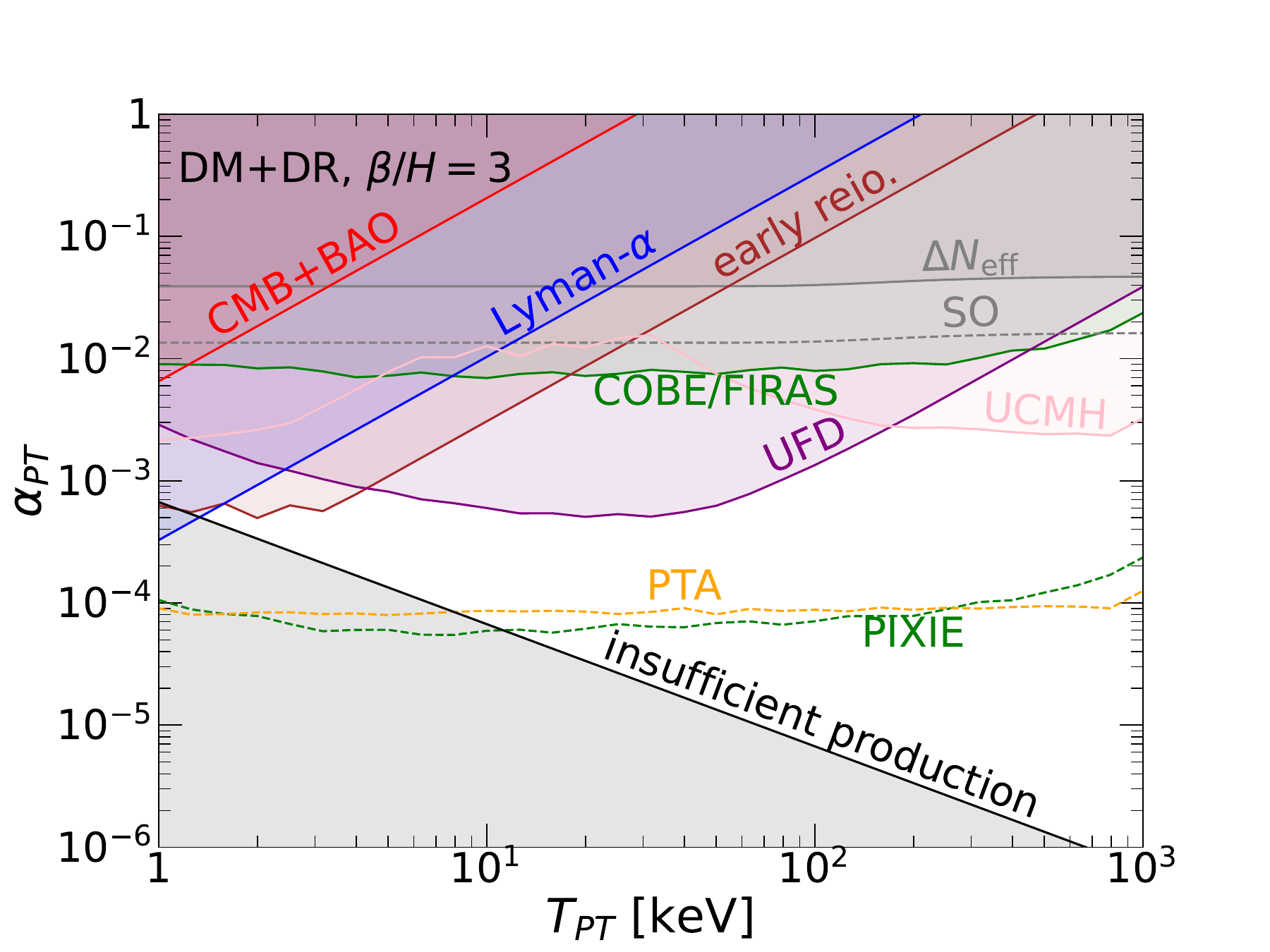}
    \caption{Bounds on the power spectrum for an optimistic benchmark $r = 1$, $\beta/H_{\rm PT} = 3$. We assume the DM+DR scenario: the latent heat is predominantly converted into DR, with the appropriate fraction being converted into cold DM to yield the correct relic abundance.
    }
    \label{fig:exclusion_broad}
\end{figure}

Our main results are presented in Figs.~\ref{fig:exclusion_zoom} and~\ref{fig:exclusion_broad}, which depict constraints on the parameter space for a super-cooled DS phase transition with $\beta/H_{\rm PT} = 3$, $T_m / T_{\rm PT} = 1$.

%%%%%%%%%%%%%%%%%%%%%%%%%%%%%%%%%%%%%%%%%%%%%%%%%%%%%%%
%%%%%%%%%%%%%%%%%%%%%%%%%%%%%%%%%%%%%%%%%%%%%%%%%%%%%%%
%%%%%%%%%%%%%%%%%%%%%%%%%%%%%%%%%%%%%%%%%%%%%%%%%%%%%%%
%%%%%%%%%%%%%%%%%%%%%%%%%%%%%%%%%%%%%%%%%%%%%%%%%%%%%%%

In Fig.~\ref{fig:exclusion_zoom}, we present constraints primarily for the DM-only scenario, in which the latent heat of the PT is predominantly converted into DM (near and below the relic-abundance curve). We zoom in on the PT temperature range $T_{\rm PT} \in (1, 10),\mathrm{keV}$, where corrections from the WDM transfer function become more significant. 

The black line shows the values of $(T_{\rm PT}, \alpha_{\rm PT})$ that reproduce the observed DM relic abundance, computed using Eq.~\eqref{eq:relicabundance}.
The strongest bounds are placed by the CMB and the Lyman-$\alpha$ forest. It turns out that the Lyman-$\alpha$ bound is stronger, although the CMB+BAO bound is more robust in the sense that our MCMC analysis allows the $\Lambda$CDM parameters to vary. For the CMB bound, we extract the $2\sigma$ boundary from Fig.~\ref{fig:MCMCtriangle} truncated to $T_{\rm PT} \in (1, 10)$~keV, which is depicted in red in Fig.~\ref{fig:exclusion_zoom}. For the Lyman-$\alpha$ forest we use the 2$\sigma$ bound from Fig.~\ref{fig:lyman}, which we show in blue in Fig.~\ref{fig:exclusion_zoom}. This excludes the minimal scenario for $T_{\rm PT} \lesssim 6$~keV.

%%%%%%%%%%%%%%%%%%%%%%%%%%%%%%%%%%%%%%%%%%%%%%%%%%%%%%%
%%%%%%%%%%%%%%%%%%%%%%%%%%%%%%%%%%%%%%%%%%%%%%%%%%%%%%%
%%%%%%%%%%%%%%%%%%%%%%%%%%%%%%%%%%%%%%%%%%%%%%%%%%%%%%%
%%%%%%%%%%%%%%%%%%%%%%%%%%%%%%%%%%%%%%%%%%%%%%%%%%%%%%%

As discussed at the beginning of Sec.~\ref{sec:analysis}, for scenarios above the relic abundance line we assume that the DS is dominated by composite matter until the perturbation modes relevant to a given cosmological constraint enter the horizon, after which the excess matter density is converted into DR. Under this assumption, the constraints in Fig.~\ref{fig:exclusion_zoom} are obtained using the enhanced initial power spectrum corresponding to the DM-like curve in Fig.~\ref{fig:powerspectrum}. They therefore represent the most aggressive constraints currently achievable for the confining DS scenario. 

A simpler way to achieve $\alpha_{\rm PT}$ above the relic abundance line is the DM+DR scenario, in which most of the latent heat of the PT is converted into DR. Thus, in Fig.~\ref{fig:exclusion_broad} we consider the DM+DR scenario and derive the CMB+BAO (red), Lyman-$\alpha$ (blue), and other constraints described in Sec.~\ref{sec:otherbounds} using the DR-like initial power spectrum illustrated in Fig.~\ref{fig:powerspectrum}, corresponding to the radiation-dominated case. Eq.~\eqref{eq:relicabundance} or (\ref{eq:minimal}) provides a lower bound on the PT strength $\alpha_{\rm PT}$, which we show in black in Fig.~\ref{fig:exclusion_broad}. Across the range of $T_{\rm PT}$ we show, the strongest constraints arise variously from the Lyman-$\alpha$ forest, early reionization (brown), heating in ultra-faint dwarfs (purple), and formation of ultracompact minihalos (pink). PIXIE (green, dashed) or a future PTA (yellow, dashed) could greatly improve upon these bounds, probing down to $\alpha_{\rm PT} = \mathcal{O}(10^{-4})$.

Comparing Figs.~\ref{fig:exclusion_zoom} and~\ref{fig:exclusion_broad}, we see that the CMB+BAO and Lyman-$\alpha$ bounds are stronger in the DM-only case than the DM+DR scenario. This is because of the enhancement of the IR tail of the power spectrum in the former scenario. Conversely, the PIXIE and PTA bounds do not show up in Fig.~\ref{fig:exclusion_zoom} because these probe the power spectrum at smaller scales, which are suppressed by warm DM effects in the DM-only scenario.

For other choices of $\beta/H_{\rm PT}$ and $v_w$, the bounds in this section can be approximately rescaled (up to growth-factor corrections) by examining the scaling of the initial curvature power spectrum in Eqs.~\eqref{eq:Pdt} and~\eqref{eq:powerspectrumfinal} with respect to the PT parameters. For scales $k/a_{\rm PT} \ll \beta$, the spectrum factorizes as
\begin{equation}\label{eq:scaling}
\mathcal{P}^{\rm DS}_\zeta(k) \sim \left[\alpha_{\rm PT}\left( \frac{\beta}{H_{\rm PT}} \right)^{-1}\right]^2 \left[v_w\left( \frac{\beta}{H_{\rm PT}} \right)^{-1}\left(k\tau_{\rm PT} \right)\right]^3\,,
\end{equation}
where $\tau_{\rm PT}\propto T_{\rm PT}^{-1}$ during radiation domination.

Consider a point in the $(T_{\rm PT}, \alpha_{\rm PT})$ plane lying on the edge of the $2\sigma$ bound. Eq.~\eqref{eq:scaling} shows that rescaling $\beta/H_{\rm PT}\rightarrow c_\beta\,\beta/H_{\rm PT}$ and taking $v_w<1$ is equivalent, at the level of the large-scale power spectrum, to shifting
\begin{equation}
\alpha_{\rm PT}' = c_\beta\,\alpha_{\rm PT}, \qquad
T_{\rm PT}' = (c_\beta v_w)^{-1} T_{\rm PT}\,.
\end{equation}
This provides a simple mapping that allows the curvature-perturbation bounds to be estimated for different choices of $\beta/H_{\rm PT}$ and $v_w$ by rescaling the $(T_{\rm PT},\alpha_{\rm PT})$ plane accordingly. Importantly, this rescaling does not affect the WDM properties. When applying this procedure to the Lyman-$\alpha$ bound in Fig.~\ref{fig:exclusion_zoom}, only the upper-right branch arising from the ``$\mathcal{P}_\zeta$ only" constraint in Fig.~\ref{fig:lyman} should be rescaled, while the region dominated by the WDM constraint should not change much.

%%%%%%%%%%%%%%%%%%%%%%%%%%
%%%%%%%%%%%%%
\section{Discussion and conclusions}\label{sec:conclusions}
It is well-motivated to consider DM production from confinement in a DS, where the relic abundance is set entirely within the DS and does not require direct couplings to the SM. In this case, the most powerful probes arise from DM's structure formation property and the cosmological perturbations sourced by the confinement dynamics, which provide a direct window into otherwise hidden DM physics. In this work, we studied the cosmological signatures of such models that undergo a strongly first-order confinement PT.

Compared to the DS phase transitions that predominantly reheat into DR and have been studied extensively in the literature, the composite-DM scenario exhibits two key differences. First, when most of the latent heat is converted into composite DM, the induced curvature power spectrum scales as $\mathcal{P}^{\rm DS}_\zeta \propto k$ on large scales, rather than the standard causal scaling $\mathcal{P}^{\rm DS}_\zeta \propto k^3$ expected for PT that reheat into radiation. Since the peak amplitude is determined by the PT duration $\beta^{-1}$ and energy-density ratio $\alpha_{\rm PT}$, this shallower scaling significantly enhances the perturbation at low-$k$ modes compared to the DR case.

Second, the composite DM is naturally produced with a sizable initial velocity, leading to warm-DM free-streaming that suppresses matter perturbations after horizon entry. The interplay between the enhanced intial curvature perturbations and the suppressed WDM transfer function gives rise to distinctive cosmological signatures.

We focused on a DM-only scenario in which the PT latent heat is predominantly converted into composite DM and accounts for the observed relic abundance. In this case, the model predicts a well-motivated relation between the PT strength $\alpha_{\rm PT}$ and the PT temperature $T_{\rm PT}$, given in Eq.~(\ref{eq:relicabundance}). For $T_{\rm PT}$ in the range $1$--$10$~keV, the composite state provides a viable DM candidate. Neglecting PT-induced perturbations, Lyman-$\alpha$ observations require $T_{\rm PT} \gtrsim 7$~keV.\footnote{This can be interpreted as a bound on the DM mass by noting $m_{\rm DM} \sim T_d(T_{\rm PT})$, c.f. Eq.~\eqref{eq:tm}. In the $1$--$10$~keV mass range, we have $\alpha_{\rm PT} \sim 10^{-4}$ and thus $m_{\rm DM} \sim 0.1 T_{\rm PT}$.}
Once PT-induced perturbations are included, they partially compensate for the suppression from free streaming, relaxing the bound to $T_{\rm PT} \gtrsim 6$~keV for the strongly supercooled PT benchmark with $\beta/H_{\rm PT}=3$. If a sizable fraction of the latent heat is instead transferred into DR, the CMB+BAO and Lyman-$\alpha$ constraints allow a stronger PT, and the compensation between the PT-induced perturbations and free-streaming suppression can further relax the bound to $T_{\rm PT} \gtrsim 5$~keV.

We remark that, since the DM abundance is related to the PT parameters, the DM model can be probed even in the absence of a nongravitational interaction with the SM. There are prospects to improve on these constraints in the future: ACT should extend CMB sensitivity to higher modes $\ell \sim 4000$ than Planck (see ACT DR6 likelihood \cite{ACT:2025fju}), and new experiments like the Simons Observatory~\cite{SO:2019} can also improve sensitivity to smaller scale perturbations in the CMB. For the Lyman-$\alpha$ forest, DESI~\cite{desiED_cat:2023,DESI_lyalpha_3D:2025,DESI_lyalpha_1DFFT:2025,DESI_lyalpha_1dOEM:2025,DESI_lyalpha_dr1valid:2026} and future experiments like New Horizon~\cite{NewHorizon:2025} could provide improved sensitivity to higher $k$-modes. 

We also studied the DM+DR scenario in which most of the latent heat is transferred into DR. In this case, the PT strength can be larger without overproducing the DM abundance, allowing constraints across a wider parameter space. In much of this region, the strongest bounds come from recently proposed probes of the power spectrum at sub-Mpc scales~\cite{Qin:2025ymc,Graham:2024hah,Bringmann:2025cht}, while a significant fraction can also be tested by future PTA experiments and PIXIE.

As a final note, bubble collisions during the PT source a stochastic gravitational wave background, providing a complementary observational signature of late-time phase transitions in addition to the matter-power-spectrum effects we study~\cite{Freese:2023fcr,Greene:2024xgq,Zebrowski:2026pye,Greene:2026one}. In particular, SKA would be sensitive to PTs above MeV, the upper end of the temperature range we consider~\cite{Schmitz:2020syl,Carilli:2004nx,Janssen:2014dka,Weltman:2018zrl}. 
This raises the intriguing possibility of discovering both the scalar and tensor perturbations arising from a PT in a confining DS.

\begin{acknowledgments}
We thank Lian-Tao Wang for helpful discussions in the early stages of this project, and Mustafa Amin and Subhajit Ghosh for insightful discussions related to this work. DH and YT are supported by the NSF Grant Number PHY-2412701. YT would also like to thank the Tom and Carolyn Marquez Chair Fund for its generous support. AI is supported by a Mafalda and Reinhard Oehme Postdoctoral Research Fellowship from the Enrico Fermi Institute at the University of Chicago. We thank the Aspen Center for Physics (supported
by NSF grant PHY-2210452), and the Munich  Institute for Astro-, Particle and BioPhysics (funded by the DFG under Germany’s Excellence Strategy EXC-2094-390783311), where this work was initiated, for their hospitality.
\end{acknowledgments}

\appendix

%%%%%%%%%%%%%%%%%%%%%%%%%%%%%%%%%%%%%%%%%%%%%%%%%%%%%%%
%%%%%%%%%%%%%%%%%%%%%%%%%%%%%%%%%%%%%%%%%%%%%%%%%%%%%%%
\section{Refined calculation of the primordial power spectrum}\label{app:transition}
%%%%%%%%%%%%%%%%%%%%%%%%%%%%%%%%%%%%%%%%%%%%%%%%%%%%%%%
%%%%%%%%%%%%%%%%%%%%%%%%%%%%%%%%%%%%%%%%%%%%%%%%%%%%%%%

We can treat the transition from radiation to matter in the DS more carefully as follows. Let us denote the average energy of a DM particle relative to its mass as $\chi = \langle E \rangle / m_{\rm DM}$. This is a function of the dark temperature $T_d$, and we can calculate it from the partition function (ignoring the integral over position, which is unimportant here):
\begin{equation}\label{eq:partition}
    Z = \int \frac{d^3 p}{(2\pi)^3} \exp \left( -\beta \sqrt{m_{\rm DM}^2 + p^2} \right) = \frac{m^3}{2\pi^2} \frac{K_2(x)}{x}
\end{equation}
where $x = m_{\rm DM} / T_d $ as usual. Then the average energy per particle is
\begin{equation}\label{eq:averageE}
    \chi = -\frac{1}{m} \frac{d \log Z}{d \beta} = \frac{3}{x} + \frac{K_1(x)}{K_2(x)} .
\end{equation}
In the ultrarelativistic (nonrelativistic) limit we recover $\chi = 3 / x$ ($\chi = 1 + 3/(2x)$), as expected.

For future convenience we note some mappings between $x$, time $t$, and wavenumber $k$. In radiation domination with $a \sim t^{1/2}$, we have $x \sim 1/T_d \sim t^{1/2}$. Let us define $t_m$ as the time when $x = 3$, corresponding to the transition from relativistic to nonrelativistic behaviour. Then $x = 3 \sqrt{t / t_m}$. A given $k$-mode enters the horizon at the time when $k = a H \sim t^{-1/2}$. This time corresponds to $x = 3 k_m / k$, where $k_m$ is the mode that enters the horizon at $t = t_m$.

The energy density in the dark sector after the PT is given by $\rho = n m_{\rm DM} \chi$. Since the number density dilutes as $a^{-3}$, it follows that the energy density scales as $a^{-3(1+w)}$, where the equation of state parameter is
\begin{equation}\label{eq:eos}
    w = -\frac{1}{3} \frac{d \log \chi}{d \log a} .
\end{equation}
By matching $\rho(t_c) = \rho_{\rm vac}$ we can derive the following expression for the energy density in the dark sector:
\begin{equation}
    \rho_d = \rho_{\rm vac} \left( \frac{t_c}{t} \right)^{3/2} \frac{\chi(x)}{\chi(x_c)} .
\end{equation}
Here $x_c$ is $x$ evaluated at $t = t_c$, which is fixed by the parameters of the PT. The density perturbation then follows as 
\begin{equation}
    \frac{\delta \rho_d}{\rho_d} = \frac{3}{2} \frac{\delta  t_c}{t_c} + \frac{d \log \chi}{d \log x} \frac{\delta x}{x} = \frac{3}{2}\left[ 1 + w(t) \right] \frac{\delta t_c}{t_c} .
\end{equation}
In the last equality we used Eq.~\eqref{eq:eos} and the fact that $x \sim a \sim 1/T_d \sim t_c^{-1/2}$.
We can then identify the curvature perturbation
\begin{equation}
    \zeta \approx - \frac{H \delta \rho_d}{\dot{\rho_{\rm SM}}} = \frac{\rho_d}{4 \rho_{\rm SM}} \frac{\delta \rho_d}{\rho_d} = \frac{\alpha_{\rm PT}}{2} \frac{\delta t_c}{t_c} \sqrt{\frac{t}{t_c}} \frac{\chi(t)}{\chi(t_c)} \frac{3}{4} \left(1 + w(t)\right) .
\end{equation}

Finally, the primordial power spectrum is given by 
\begin{equation}
    \mathcal{P}_\zeta(k) = \alpha_{\rm PT}^2 \mathcal{P}_{\delta t}(k) \left[ \frac{3}{4} \left(1 + w(k) \right) \right]^2 F(k)^2, \quad F(k) = \frac{k_c}{k} \frac{\chi(k)}{\chi(k_c)} .
\end{equation}
We defined $k_c$ as the mode that enters the horizon at $t = t_c$; $w(k)$ refers to $w$ evaluated at the time that the mode $k$ enters the horizon.

It is easy to see that this reproduces Eq.~\eqref{eq:powerspectrumfinal} for large and small $k$. Assume that the DS is initially relativistic, so we can approximate $\chi(k_c) \approx 3 / x = k_c / k_m$ and $w(k_c) = 1/3$. For modes with $k \gg k_m$, which enter when the DS is still relativistic, we can use the same approximation $\chi(k) \approx k / k_m$. It follows that $F(k) \approx 1$ in this regime. For $k \ll k_m$ the mode enters when the DS is nonrelativistic, so $\chi \approx 1$ and $w(k) \approx 0$. Then we have $F(k) = k_m / k$, leading to a $1/k^2$ enhancement for such modes.

%%%%%%%%%%%%%%%%%%%%%%%%%%%%%%%%%%%%%%%%%%%%%%%%%%%%%%%
%%%%%%%%%%%%%%%%%%%%%%%%%%%%%%%%%%%%%%%%%%%%%%%%%%%%%%%
\section{Superhorizon evolution equation}\label{app:superhorizon}
%%%%%%%%%%%%%%%%%%%%%%%%%%%%%%%%%%%%%%%%%%%%%%%%%%%%%%%
%%%%%%%%%%%%%%%%%%%%%%%%%%%%%%%%%%%%%%%%%%%%%%%%%%%%%%%

In this appendix, we corroborate our result for the curvature perturbation Eq.~\eqref{eq:curvature} derived in the main text with the superhorizon evolution equation. Our starting point is the equation for the time evolution of curvature perturbations on large scales~\cite{Wands:2000dp}:
\begin{equation}\label{eq:wands}
    \dot{\zeta} = -\frac{H}{p+\rho} \left[\delta p - \frac{\dot{p}}{\dot{\rho}}\delta\rho \right] ,
\end{equation}
where $p$ and $\rho$ are the total pressure and energy density respectively. We consider the DM-only scenario: all of the latent heat $\rho_{\rm vac}$ of the PT is \textit{instantaneously} converted into ultrarelativistic DM composites at $t_c$, and the DM composites \textit{instantaneously} become nonrelativistic at some later time $t_m > t_c$.     
The approximation that these changes happen instantaneously will lead to a discontinuity in the curvature perturbation; a more careful, continuous treatment of the transition at $t_m$ was given in Appendix~\ref{app:transition}. For the following derivation, we distinguish the space-averaged transition times $\Bar{t}_c$ and $\Bar{t}_m$ from the spatial fluctuations $\delta t_c = t_c - \Bar{t}_c$ and $\delta t_m = t_m - \Bar{t}_m$. We assume that these perturbations are small $\delta t_c/\Bar{t}_c,\, \delta t_m/\Bar{t}_m \ll 1$ such that $\Bar{t}_c$ and $\Bar{t}_m$ serve as approximate reference times for when the respective transitions are completed across all regions of space.  

The \textit{local} energy density in the confining DS is described by
\begin{align}
    \rho_{d}(t) = 
    \begin{cases}
        \rho_{\rm vac} \hspace{40mm}t<t_c\\
        \rho_{\rm vac}\left( \frac{t}{t_c} \right)^{-2} \hspace{27mm} t_c<t<t_m\\
        \rho_{\rm vac}\left( \frac{t_m}{t_c} \right)^{-2}\left( \frac{t}{t_m} \right)^{-3/2} \hspace{10mm} t>t_m
    \end{cases} ,
\end{align}
with the DM composites redshifting as radiation for $t<t_m$ and as matter for $t>t_m$. The corresponding DS equation of state is
\begin{align}
    \omega_{d}(t) = 
    \begin{cases}
        -1  \hspace{21mm}t<t_c\\
        1/3 \hspace{20mm} t_c<t<t_m\\
        0   \hspace{24mm} t>t_m
    \end{cases} .
\end{align}
In the radiation era, the remaining energy density of the universe will be dominated by the SM radiation. Using $\Bar{t}_c$ as a reference time, this is given by
\begin{align}
    \rho_{\rm SM} = \rho_{\rm SM}(T_{\rm PT}) \left( \frac{t}{\Bar{t}_c} \right)^{-2} 
\end{align}
with equation of state $\omega_{\rm SM} = 1/3$. The total energy density $\rho =\sum_i\rho_i$ is then 
\begin{equation}\begin{split}
    \rho &=\ \rho_{\rm vac} \Theta(t_c - t)\ +\ \rho_{\rm vac}\left( \frac{t}{t_c} \right)^{-2}\Theta(t - t_c)\Theta(t_m - t)\ \\
    &+\ \rho_{\rm vac}\left( \frac{t_m}{t_c} \right)^{-2}\left( \frac{t}{t_m} \right)^{-3/2}\Theta(t - t_m)\ +\ \rho_{\rm SM}(T_{\rm PT}) \left( \frac{t}{\Bar{t}_c} \right)^{-2}
\end{split}\end{equation}
and the total pressure $p =\sum_i\omega_i\rho_i$ is
\begin{equation}
    \hspace{-5mm}p =\ -\rho_{\rm vac} \Theta(t_c - t)\ +\ \frac{1}{3}\rho_{\rm vac}\left( \frac{t}{t_c} \right)^{-2}\Theta(t - t_c)\Theta(t_m - t)\ +\ \frac{1}{3}\rho_{\rm SM}(T_{\rm PT}) \left( \frac{t}{\Bar{t}_c} \right)^{-2} .
\end{equation}
From these expressions we can compute the time derivatives of the background quantities:
\begin{align}
    \hspace{-7mm}\dot{\rho} = &-4H\rho_{\rm vac}\left( \frac{t}{\Bar{t}_c} \right)^{-2}\Theta(t - \Bar{t}_c)\Theta(\Bar{t}_m-t)\notag\\ 
    &\hspace{10mm}-\ 3H\rho_{\rm vac}\left( \frac{\Bar{t}_m}{\Bar{t}_c} \right)^{-2}\left( \frac{t}{\Bar{t}_m} \right)^{-3/2}\Theta(t-\Bar{t}_m)\ -\ 4H\rho_{\rm SM}(T_{\rm PT}) \left( \frac{t}{\Bar{t}_c} \right)^{-2}\notag\\
    =&-\ 4H\rho_{\rm SM}(T_{\rm PT}) \left( \frac{t}{\Bar{t}_c} \right)^{-2}\left\{ 1 + \alpha_{\rm PT}\Theta(t - \Bar{t}_c)\Theta(\Bar{t}_m-t) + \frac{3}{4}\alpha_{\rm PT}\sqrt{\frac{t}{\Bar{t}_m}}\Theta(t-\Bar{t}_m)\right\}\,,
\end{align}
and
\begin{align}
    \hspace{-10mm}\dot{p} = &\ \frac{4}{3}\rho_{\rm vac} \delta(t - \Bar{t}_c)\ -\ \frac{4}{3}H\rho_{\rm vac}\left( \frac{t}{\Bar{t}_c}\right)^{-2}\Theta(t-\Bar{t}_c)\Theta(\Bar{t}_m-t)\notag\\ 
    &\hspace{10mm}-\ \frac{1}{3}\rho_{\rm vac}\left( \frac{t}{\Bar{t}_c}\right)^{-2}\delta(t-\Bar{t}_m) -\ \frac{4}{3}H\rho_{\rm SM}(T_{\rm PT}) \left( \frac{t}{\Bar{t}_c} \right)^{-2}\notag\\
    =&-\ \frac{4}{3}H\rho_{\rm SM}(T_{\rm PT}) \left( \frac{t}{\Bar{t}_c} \right)^{-2}\left\{ 1-\frac{1}{H_{\rm PT}}\alpha_{\rm PT}\delta(t-\Bar{t}_c)+\alpha_{\rm PT}\Theta(t-\Bar{t}_c)\Theta(\Bar{t}_m-t) +\frac{1}{4H_m}\alpha_{\rm PT}\delta(t-\Bar{t}_m)\right\}\,,
\end{align}
using $H =  1/2t$ in the radiation era. 
We also set $t=\Bar{t}_c$ or $t=\Bar{t}_m$ under the respective delta functions and denote $H_{\rm PT}\equiv 1/2\Bar{t}_c$ and $H_m \equiv 1/2\Bar{t}_m$. Likewise, the perturbations can be obtained by varying with respect to $t_c$ and $t_m$ to linear order, noting that $\delta t_m/\Bar{t}_m = \delta t_c/\Bar{t}_c$ since $t_c \propto t_m$: 
\begin{align}
    \delta\rho =&\ 4\rho_{\rm vac}\left( \frac{t}{\Bar{t}_c}\right)^{-2}\Theta(t-\Bar{t}_c)\Theta(\Bar{t}_m-t)\frac{\delta t_c}{2\Bar{t}_c}\ +\ 3\rho_{\rm vac}\left( \frac{\Bar{t}_m}{\Bar{t}_c}\right)^{-2}\left( \frac{t}{\Bar{t}_m} \right)^{-3/2}\Theta(t-\Bar{t}_m)\frac{\delta t_m}{2\Bar{t}_m}\notag\\
    =&\ 4\rho_{\rm vac}\left( \frac{t}{\Bar{t}_c} \right)^{-2}\left\{\Theta(t-\Bar{t}_c)\Theta(\Bar{t}_m-t)\ +\ \frac{3}{4}\sqrt{\frac{t}{\Bar{t}_m}}\Theta(t-\Bar{t}_m) \right\}\frac{\delta t_c}{2\Bar{t}_c}\,,
\end{align}
and
\begin{align}
    \hspace{-10mm}\delta p =&\ -\frac{4}{3}\rho_{\rm vac}\delta(t-\Bar{t}_c)\delta t_c\ +\ \frac{4}{3}\rho_{\rm vac}\left( \frac{t}{\Bar{t}_c}\right)^{-2}\Theta(t-\Bar{t}_c)\Theta(\Bar{t}_m-t)\frac{\delta t_c}{2\Bar{t}_c}\ +\ \frac{1}{3}\rho_{\rm vac}\left( \frac{t}{\Bar{t}_c}\right)^{-2}\delta(t-\Bar{t}_m)\delta t_m\notag\\
    =&\ \frac{4}{3}\rho_{\rm vac}\left( \frac{t}{\Bar{t}_c} \right)^{-2}\left\{-\frac{1}{H_{\rm PT}}\delta(t-\Bar{t}_c)\ +\ \Theta(t-\Bar{t}_c)\Theta(\Bar{t}_m-t)\ +\ \frac{1}{4H_m}\delta(t-\Bar{t}_m)\right\}\frac{\delta t_c}{2\Bar{t}_c}\,.
\end{align}
With these expressions, we can find
\begin{equation}
    \frac{\dot{p}}{\dot{\rho}} =\ \frac{1}{3}\ \frac{1 + \alpha_{\rm PT}\left[-\frac{1}{H_{\rm PT}}\delta(t-\Bar{t}_c) + \frac{1}{4H_m}\delta(t-\Bar{t}_m) + \Theta(t-\Bar{t}_c)\Theta(\Bar{t}_m-t)\right]}{1 + \alpha_{\rm PT}\Theta(t-\Bar{t}_c)\Theta(\Bar{t}_m-t) + \frac{3}{4}\alpha_{\rm PT}\sqrt{\frac{t}{\Bar{t}_m}}\Theta(t-\Bar{t}_m)}\,,
\end{equation}
\begin{align}
    \hspace{-5mm}\frac{\dot{p}}{\dot{\rho}}\delta\rho =\ \frac{4}{3}\rho_{\rm vac}\left( \frac{t}{\Bar{t}_c} \right)^{-2}\Bigg\{& \frac{\alpha_{\rm PT}\Theta(t-\Bar{t}_c)\Theta(\Bar{t}_m-t) + \frac{3}{4}\alpha_{\rm PT}\sqrt{\frac{t}{\Bar{t}_m}}\Theta(t-\Bar{t}_m)}{1 + \alpha_{\rm PT}\Theta(t-\Bar{t}_c)\Theta(\Bar{t}_m-t) + \frac{3}{4}\alpha_{\rm PT}\sqrt{\frac{t}{\Bar{t}_m}}\Theta(t-\Bar{t}_m)}\notag\\
    &\times \Bigg[-\frac{1}{H_{\rm PT}}\delta(t-\Bar{t}_c)\ +\ \frac{1}{4H_m}\delta(t-\Bar{t}_m)\ +\ \Theta(t-\Bar{t}_c)\Theta(\Bar{t}_m-t) \Bigg]\notag\\
     &\hspace{15mm}+\frac{\Theta(t-\Bar{t}_c)\Theta(\Bar{t}_m-t) + \frac{3}{4}\sqrt{\frac{t}{\Bar{t}_m}}\Theta(t-\Bar{t}_m)}{1 + \alpha_{\rm PT}\Theta(t-\Bar{t}_c)\Theta(\Bar{t}_m-t) + \frac{3}{4}\alpha_{\rm PT}\sqrt{\frac{t}{\Bar{t}_m}}\Theta(t-\Bar{t}_m)}\Bigg\}\frac{\delta t_c}{2\Bar{t}_c}\,,
\end{align}
\begin{align}
    \hspace{-10mm}\delta p -\frac{\dot{p}}{\dot{\rho}}\delta\rho &=\ \frac{4}{3}\rho_{\rm vac}\left( \frac{t}{\Bar{t}_c} \right)^{-2}\Bigg\{ \Bigg[1 - \frac{\alpha_{\rm PT}\Theta(t-\Bar{t}_c)\Theta(\Bar{t}_m-t) + \frac{3}{4}\alpha_{\rm PT}\sqrt{\frac{t}{\Bar{t}_m}}\Theta(t-\Bar{t}_m)}{1 + \alpha_{\rm PT}\Theta(t-\Bar{t}_c)\Theta(\Bar{t}_m-t) + \frac{3}{4}\alpha_{\rm PT}\sqrt{\frac{t}{\Bar{t}_m}}\Theta(t-\Bar{t}_m)}\Bigg]\notag\\
    &\hspace{35mm}\times \Bigg[-\frac{1}{H_{\rm PT}}\delta(t-\Bar{t}_c)\ +\ \frac{1}{4H_m}\delta(t-\Bar{t}_m)\ +\ \Theta(t-\Bar{t}_c)\Theta(\Bar{t}_m-t) \Bigg]\notag\\
     &\hspace{45mm}-\frac{\Theta(t-\Bar{t}_c)\Theta(\Bar{t}_m-t) + \frac{3}{4}\sqrt{\frac{t}{\Bar{t}_m}}\Theta(t-\Bar{t}_m)}{1 + \alpha_{\rm PT}\Theta(t-\Bar{t}_c)\Theta(\Bar{t}_m-t) + \frac{3}{4}\alpha_{\rm PT}\sqrt{\frac{t}{\Bar{t}_m}}\Theta(t-\Bar{t}_m)}\Bigg\}\frac{\delta t_c}{2\Bar{t}_c}\notag\\
     &=\ \frac{4}{3}\rho_{\rm vac}\left( \frac{t}{\Bar{t}_c} \right)^{-2}\Bigg\{ \frac{-\frac{1}{H_{\rm PT}}\delta(t-\Bar{t}_c)\ +\ \frac{1}{4H_m}\delta(t-\Bar{t}_m) - \frac{3}{4}\sqrt{\frac{t}{\Bar{t}_m}}\Theta(t-\Bar{t}_m)}{1 + \alpha_{\rm PT}\Theta(t-\Bar{t}_c)\Theta(\Bar{t}_m-t) + \frac{3}{4}\alpha_{\rm PT}\sqrt{\frac{t}{\Bar{t}_m}}\Theta(t-\Bar{t}_m)}\Bigg\}\frac{\delta t_c}{2\Bar{t}_c}\,.
\end{align}
We can then evaluate the source term of Eq.~\eqref{eq:wands}: 
\begin{equation}
\dot{\zeta} =\ \alpha_{\rm PT}\left\{ \frac{\delta(t-\Bar{t}_c)\ -\ \frac{1}{4}\delta(t-\Bar{t}_m) + \frac{3}{4}H\sqrt{\frac{t}{\Bar{t}_m}}\Theta(t-\Bar{t}_m)}{\left(1 + \alpha_{\rm PT}\Theta(t-\Bar{t}_c)\Theta(\Bar{t}_m-t) + \frac{3}{4}\alpha_{\rm PT}\sqrt{\frac{t}{\Bar{t}_m}}\Theta(t-\Bar{t}_m)\right)^2}\right\}\frac{\delta t_c}{2\Bar{t}_c}\,.   
\end{equation}
Working to leading order in $\alpha_{PT} \ll 1$, this simplifies to: 
\begin{equation}
\dot{\zeta} \approx\ \alpha_{\rm PT}\left\{\delta(t-\Bar{t}_c)\ -\ \frac{1}{4}\delta(t-\Bar{t}_m) + \frac{3}{4}\left(\frac{d}{dt}\sqrt{\frac{t}{\Bar{t}_m}}\right)\Theta(t-\Bar{t}_m)\right\}\frac{\delta t_c}{2\Bar{t}_c}\ +\ {\cal O}(\alpha_{\rm PT}^2) \,,     
\end{equation}
noting that $H\sqrt{t/\Bar{t}_m}$ is simply the time derivative of $\sqrt{t/\Bar{t}_m}$. The curvature perturbation generated by the PT can be obtained by integrating this to some later time $t>\Bar{t}_c$ after PT completion. Let us denote the integral of the terms in the curly brackets by
\begin{align}
    F(t) &\equiv \int_0^t dt'\left\{\delta(t'-\Bar{t}_c)\ -\ \frac{1}{4}\delta(t'-\Bar{t}_m) + \frac{3}{4}\left(\frac{d}{dt'}\sqrt{\frac{t'}{\Bar{t}_m}}\right)\Theta(t' - \Bar{t}_m)\right\}\,.
\end{align}
The last two terms in the integral vanishes if $t<\Bar{t}_m$ such that 
\begin{equation}
    F(t<\Bar{t}_m) = \int_0^t dt'\delta(t'-\Bar{t}_c) = 1\,,
\end{equation}
but if $t>\Bar{t}_m$, we obtain:
\begin{equation}
    F(t>\Bar{t}_m) =\ 1 -\ \frac{1}{4}\ +\ \int_{\Bar{t}_m}^t dt'\ \frac{3}{4}\left(\frac{d}{dt'}\sqrt{\frac{t'}{\Bar{t}_m}}\right) = \frac{3}{4}\sqrt{\frac{t}{\Bar{t}_m}}\,.
\end{equation}
All together, the curvature perturbation is
\begin{align}
    \zeta &= \alpha_{\rm PT}F(t)\frac{\delta t_c}{2\Bar{t}_c}\,,\\
    F(t) &=
    \begin{cases}
        1\ \qquad\qquad\qquad t<\Bar{t}_m\\
        \frac{3}{4}\sqrt{\frac{t}{\Bar{t}_m}}\qquad\qquad t>\Bar{t}_m\,.
    \end{cases}
\end{align}
as found in Eq.~\eqref{eq:curvature}. Notably, the amplitude changes by a factor of $3/4$ when the DM composites become non-relativistic at $\Bar{t}_m$ and an isocurvature fluctuation is generated between radiation ($\omega_d = 1/3$) and matter ($\omega_d = 0$). Since $t_m \propto t_c$, the curvature perturbation sourced by this $\delta t_m$-fluctuation has an opposite sign compared to the initial curvature perturbation sourced by the $\delta t_c$-fluctuation at PT. Just as patches which underwent PT earlier redshifted \textit{faster} as radiation relative to the false vacuum, the same patches now become nonrelativistic earlier and redshift \textit{slower} as matter relative to radiation, leading to some cancellation in the final curvature perturbation.

\bibliographystyle{JHEP}
\bibliography{references}

\end{document}